\documentclass{aastex62}
\usepackage{graphicx}	
\usepackage{amsmath}	
\usepackage{amssymb}	
\usepackage{natbib}     
\usepackage{subfigure}
\usepackage{booktabs}
\usepackage{tabularx}
\usepackage{mathrsfs}

	\graphicspath{{./}{figures/}}

	\received{}
	\revised{}
	\accepted{}
	\submitjournal{ApJ}
	
	\shorttitle{PSR~B0943$+$10}
	\shortauthors{Cao et al.}
	
\begin{document}
			
			\title{
				PSR~B0943$+$10: Mode Switch, Polar Cap Geometry, and Orthogonally Polarized Radiation
			}%

			\correspondingauthor{Jinchen Jiang, Kejia Lee, Renxin Xu}
			\email{jiangjinchen@bao.ac.cn, kjlee@pku.edu.cn, r.x.xu@pku.edu.cn}
			
			\author{Shunshun Cao}
			\affil{Department of Astronomy, School of Physics, Peking University, Beijing 100871, China}
			
			\author{Jinchen Jiang}
			\affiliation{National Astronomical Observatories, Chinese Academy of Sciences, Beijing 100012, China}
			
			\author{Jaroslaw Dyks}
			\affiliation{Nicolaus Copernicus Astronomical Center, Polish Academy of Sciences, Rabia\'{n}ska 8, 87-100, Toru\'{n}, Poland}
			
			\author{Longfei Hao}
			\affiliation{Yunnan Observatories, Chinese Academy of Sciences, 650216 Kunming, China}
			\affiliation{Key Laboratory for the Structure and Evolution of Celestial Objects, Chinese Academy of Sciences, 650216 Kunming, China}
			
			\author{Kejia Lee}
			\affiliation{Department of Astronomy, School of Physics,
				Peking University, Beijing 100871, China}
			\affiliation{Kavli Institute for Astronomy and
				Astrophysics, Peking University, Beijing 100871, China}
			\affiliation{National Astronomical Observatories, Chinese Academy of Sciences, Beijing 100012, China}
			
			\author{Zhixuan Li}
			\affiliation{Yunnan Observatories, Chinese Academy of Sciences, 650216 Kunming, China}
			\affiliation{Key Laboratory for the Structure and Evolution of Celestial Objects, Chinese Academy of Sciences, 650216 Kunming, China}
			
			\author{Jiguang Lu}
			\affiliation{National Astronomical Observatories, Chinese Academy of Sciences, Beijing 100012, China}
			\affiliation{Guizhou Radio Astronomical Observatory, Guiyang 550025, China}
			
			\author{Zhichen Pan}
			\affiliation{National Astronomical Observatories, Chinese Academy of Sciences, Beijing 100012, China}
			\affiliation{CAS Key Laboratory of FAST, National Astronomical Observatories, Chinese Academy of Sciences, Beijing 100101, China}
			\affiliation{School of Astronomy and Space Science, University of Chinese Academy of Sciences, Beijing 100049, China}
			
			\author{Weiyang Wang}
			\affiliation{School of Astronomy and Space Science, University of Chinese Academy of Sciences, Beijing 100049, China}
			
			\author{Zhengli Wang}
			\affiliation{Guangxi Key Laboratory for Relativistic Astrophysics, School of Physical Science and Technology, Guangxi University, Nanning 530004, China}
			
			\author{Jiangwei Xu}
			\affiliation{Department of Astronomy, School of Physics,
				Peking University, Beijing 100871, China}
			\affiliation{Kavli Institute for Astronomy and
				Astrophysics, Peking University, Beijing 100871, China}
			\affiliation{National Astronomical Observatories, Chinese Academy of Sciences, Beijing 100012, China}
			
			\author{Heng Xu}
			\affiliation{National Astronomical Observatories, Chinese Academy of Sciences, Beijing 100012, China}
			
			\author{Renxin Xu}
			\affiliation{Department of Astronomy, School of Physics,
				Peking University, Beijing 100871, China}
			\affiliation{Kavli Institute for Astronomy and
				Astrophysics, Peking University, Beijing 100871, China}

			
			
			\begin{abstract}
				As one of the paradigm examples to probe into pulsar magnetospheric dynamics, PSR B0943$+$10 (J0946$+$0951) manifests representatively, showing mode switch, orthogonal polarization and subpulse drifting, frequently studied below 600~MHz.
				Here both integrated and single pulses are studied at a high frequency (1.25~GHz) with FAST.
				%
				The mode switch is studied using a profile decomposition method. A phase space evolution for the pulsar's mode switch shows a strange-attractor-like pattern.
				The radiative geometry is proposed by fitting polarization position angles with the rotating vector model. The pulsar pulse profile is then mapped to the sparking locations on pulsar surface, and the differences between the main pulse's and the precursor component's radiative processes may explain the X-ray's synchronization with radio mode switch.
				Detailed single pulse studies on B0943$+$10's orthogonally polarized radiation are presented, which may support for certain models of radiative transfer of polarized emission. Especially, the difference in OPMs' circular polarization might reflect the cyclotron absorption in pulsar magnetospheres.
				B0943$+$10's B and Q modes evolve differently with frequency and have different proportions of orthogonal modes, which indicates possible magnetospheric changes during mode switch. For Q mode pulse profile, the precursor and the main pulse components are orthogonally polarized, and are probably originated from different depths in the magnetosphere.
				The findings could impact significantly on pulsar electrodynamics and the radiative mechanism related.
			\end{abstract}
			
		
		
		\section{Introduction} \label{sec:intro}
		%
		Pulsars are usually interpreted as highly magnetized rotating compact stars emitting pulse-like periodic signals. Their abundant radiation properties are closely related to some basic problems in pulsar physics, but are still poorly understood. PSR~B0943$+$10 (J0946$+$0951) is a well-known pulsar rich in radiation phenomena. It is a typical mode-switching pulsar with two stable radiation modes showing distinct radiation phenomena. There exists a B (``Burst'' or ``Bright'') mode with organized simple sub-pulse drifting pattern and a Q (``Quiescent'' or ``Quiet'') mode with disorganized single pulses~\citep[e.g.,][]{1998JApA...19....1S,deshpande2001}. Since its discovery in 1968~\citep{vitkevich1969}, many radio observations have been carried out on B0943$+$10~\citep[e.g.,][]{1998JApA...19....1S,deshpande1999,2011MNRAS.418.1736B,2014A&A...572A..52B}. Radio observations on B0943$+$10 are mostly made on relatively low frequencies (below 600~MHz), because the source becomes dimmer quickly when frequency increases~\citep{deshpande1999} (also see Arecibo L band observation of B0943$+$10 in~\cite{1999ApJS..121..171W}).
		
		\cite{1998JApA...19....1S} and~\cite{deshpande2001} studied B0943$+$10's orthogonal polarization modes (OPMs). They introduced a primary polarization mode (PPM), a secondary polarization mode (SPM) and a unpolarized mode (UP) (see Figure~16 in~\cite{deshpande2001}) to describe its polarized pulse sequence. PPM and SPM are orthogonally polarized. It was found that B0943$+$10's radiation modes have different proportions of PPM and SPM at 103~MHz~\citep{1998JApA...19....1S}: B mode is PPM dominated (90\%) while Q mode is SPM dominated (59\%), leading to Q mode profile's lower linear polarization fraction. Moreover,~\cite{deshpande2001} applied a polarization-mode separation technique to B0943$+$10's B mode profiles at 430~MHz, and found that the derived PPM and SPM profiles show opposite senses of circular polarization: PPM has left-hand circularity (Stokes $V>0$) and SPM has right-hand circularity (Stokes $V<0$). In the last 20 years, attentions have mostly been paid on this pulsar's sub-pulse drifting phenomena and profile evolution~\citep[e.g.,][]{2006A&A...453..679R, 2009MNRAS.396..870S, 2014A&A...572A..52B, 2021MNRAS.502.6094S}. Now with the help of China's Five-hundred-meter Aperture Spherical radio Telescope (FAST), more studies on B0943$+$10's single pulses and polarization can be made, under a relatively high frequency: 1-1.5~GHz. We haven't made new progress on B0943$+$10's sub-pulse drifting, and this paper concentrates on mode switch and polarization properties.
		
		Beside radio observations, X-ray observations of B0943$+$10 raise more questions. \cite{2013Sci...339..436H} found that B0943$+$10 shows different ways of X-ray emission in different radio radiation modes: a thermal X-ray pulsation is strong in Q mode, but weak in B mode. Further analysis done by \cite{2016ApJ...831...21M} showed that Q mode X-ray emission is well described with a blackbody pulsed component and a power law unpulsed component, and such kind of components' combination could also explain the B mode X-ray emission. However, the frequently mentioned nearly-aligned-rotator geometry given by \cite{deshpande2001} is a barrier for understanding B0943$+$10's X-ray pulsation simply from polar cap hotspots. The underlying mechanism of the relation between X-ray emission and radio emission modes remains largely unknown.
  
        In this paper, we perform detailed studies towards understanding B0943$+$10's radio and X-ray emission, based on FAST data. Information on our observation data and processing is given in Section~\ref{sec:sec2}. To quantitatively describe the whole mode switch process, we apply a decomposition method to the pulse profiles, and the algorithm is introduced in this section. Besides, the method of mapping the radiation to the polar cap region in \cite{Wang_2024} is briefly described in this section, too.
        
        \par Section~\ref{sec:sec3} is divided into three parts. The first part is about the mode evolution in time domain analyzed through the decomposition algorithm in Section~\ref{sec:sec2}, as well as the frequency evolution of profile: under FAST L band, Q mode is found to be almost as bright as B mode. Full polarization analysis on integrated profiles and single pulses are carried out in the second and the third part. After taking into account the differently dominating OPMs for the precursor component and the main pulse component, B0943$+$10's radiation well follows the rotating vector model~\citep[RVM,][]{RVM1969}. And the result is different from that of \cite{deshpande2001}. With the radiation geometry, the possible surface origin of radiation particles, and estimations of radiation particles' Lorentz factors and heights, could be derived with the geometric mapping method in~\cite{Wang_2024}. The OPMs' relation with single pulses' circular polarization and the OPMs' proportions in different radiation modes are also included in this section.

        \par Section~\ref{sec:sec4} contains various discussions on the results. One of the most challenging topics is to understand the X-ray pulsation properties mentioned above. Firstly, for the oblique radiation geometry obtained in Section~\ref{sec:sec3}, the X-ray pulsation in both modes are easy to appear when there is a hotspot in the polar cap region. Besides, we argue that emission heights of different pulse components, and even of different radiation modes, might be significantly different, which means that the Lorentz factor of the radiating particles could be different for B and Q modes. Then the difference between flowing back particles' energy leads to radiation modes' difference in the temperature of hotspots, which could account for the X-ray pulsation intensity's variation in B and Q modes. 
        
        \par Discussions are also made on the OPM properties. The difference in single pulse samples' circular polarization distributions for OPMs Pulse samples dominated by different orthogonal modes tend to have circular polarization of opposite handedness, which could be related to propagation effects especially the mode coupling and cyclotron absorption in the pulsar magnetosphere~\citep[e.g.,][]{2001A&A...378..883P, 2010MNRAS.403..569W,2012MNRAS.425..814B}, according to which we analyze the possible specific processes of wave propagation in B0943$+$10's magnetosphere. With the above arguments, we discuss on B0943$+$10's mode switch trigger at the end of Section~\ref{sec:sec4}. The conclusions of this paper are given in Section~\ref{sec:sec5}.
		
		\section{Observation and data reduction}\label{sec:sec2}
		Four epochs of observation during May, 2022 to August, 2023 are used in this paper. All of them are made by FAST at small zenith angles ($ < 26.4^{\circ}$), with L-band 19-beam receiver \citep{2020RAA....20...64J}. The data is recorded under the frequency band 1000~MHz to 1500~MHz, and the band is divided into 4096 channels. The time resolution of the recording is 49.152$\mu$s. At the beginning of each observation, modulated signals from a noise diode were injected as a 100\% linearly polarized source for polarimetric calibration. The data processing (including folding, RFI mitigation, calibration and timing) is done with software packages \textsc{dspsr} \citep{2011PASA...28....1V}, \textsc{psrchive} \citep{2004PASA...21..302H} and \textsc{tempo2} \citep{2006MNRAS.369..655H}. Each pulse period is divided into 1024 bins. In this paper, we follow the PSR/IEEE convention for the definition of Stokes parameters \citep{2010PASA...27..104V}. When timing the time-of-arrivals, considering that B and Q modes' profiles can be significantly different, we make two templates aligned in one observation separately for B and Q mode within epoch (b), to calculate the time-of-arrival for 4 epochs. We use a Bayesian method to fit for the rotation measure (RM) of Faraday rotation \citep{2020Natur.586..693L} from the calibrated data. For single pulses, baselines of system noise are fitted with linear functions and then subtracted. Table~\ref{tab:par} shows basic information of 4 epochs and the updated RM values using FAST data. The RM changes shown in Table~\ref{tab:par} are probably due to variations of the earth's ionosphere, so we estimate the ionospheric contribution using software \texttt{ionFR} \citep{2013A&A...552A..58S,2013ascl.soft03022S} with the global ionosphere map (GIM) UPCG\footnote{https://cddis.nasa.gov/archive/gnss/products/ionex/} \citep{1999JASTP..61.1237H}. 
		
		\begin{table}
			\centering
			\renewcommand\arraystretch{1.2}
			\caption{SOME PARAMETERS OF PSR~B0943$+$10. $\mathrm{RM_{obs}}$ is the rotation measure fitted using observation data. $\mathrm{RM_{ion}}$ is the ionospheric rotation measure estimated using software \texttt{ionFR} with the global ionosphere map (GIM) UPCG, and the RM contributed by the interstellar medium (ISM) $\mathrm{RM_{ISM}}=\mathrm{RM_{obs}}-\mathrm{RM_{ion}}$. \label{tab:par}}
			
			\begin{tabular}{ccccc} 
				\hline
				\hline
				Observation date (UTC) & Time length (s) & $\mathrm{RM_{obs}}$ ($\mathrm{rad\,m^{-2}}$) & $\mathrm{RM_{ion}}$ ($\mathrm{rad\,m^{-2}}$) & $\mathrm{RM_{ISM}}$ ($\mathrm{rad\,m^{-2}}$)\\
				\hline
				20220517 (a) ................... & 7335 & 17.3$\pm0.5$ & $1.69\pm 0.08$ & $15.6\pm 0.5$ \\
				20220902 (b) ................... & 3000 & 18.7$\pm0.4$ & $3.64\pm 0.08$ & $15.0\pm 0.4$ \\
				20230816 (c) ................... & 3280 & 20.1$\pm0.4$ & $5.68\pm 0.08$ & $14.4\pm 0.3$\\
				20230827 (d) ................... & 2720 & 19.1$\pm0.4$ & $6.04\pm 0.08$ & $13.1\pm 0.3$\\
				\hline
			\end{tabular}
		\end{table}

		\subsection{Profile decomposition: eigen mode search}\label{sec:modesearch}
		
		To make a clear description of the mode switches and other profile evolution processes,~\citet[][submitted]{2023Hao_submitted} put forward an eigen mode searching method. We extend the method from only calculating eigen modes of $I$ to calculating eigen modes of polarized profiles $I$, $L$ and $V$ together. Any sub-integrations (noted with $p_{ij}$, $i$-th sub-integration's $j$-th bin) could be linearly decomposed into some eigen modes ($f_{j}$ and $g_{j}$, take two modes for example) together with a certain amount of noise ($n_{ij}$): 
		\begin{equation}
			p_{ij}=\alpha_{i}f_{j} +\beta_{i}g_{j}+n_{ij}
			\label{equa:pij}
		\end{equation}
		Here $\alpha_{i}$ and $\beta_{i}$ represent the mixture weights of eigen modes $f_{j}$ and $g_{j}$. Assuming $n_{ij}$ follows Gaussian distribution, the likelihood used to estimate the modes and the weights could be written in the form:
		\begin{equation}
			\Lambda \propto e^{-\dfrac{1}{2}\sum_{i}\sum_{j}\left(\dfrac{p_{ij}-\alpha_{i}f_{j}-\beta_{i}g_{j}}{\sigma_{ij}}\right)^{2}}
			\label{equa:likelihood}
		\end{equation}
		Here $\sigma_{ij}$ means the uncertainty of the sub-integration profile value at the $i$-th sub-integration profile's $j$-th bin. The relations between the best estimated parameters are derived from $\partial \Lambda /\partial \alpha_{i}=0$, $\partial \Lambda /\partial \beta_{i}=0$, $\partial \Lambda /\partial f_{j}=0$ and $\partial \Lambda /\partial g_{j}=0$:
		\begin{equation}
			\sum_{j}(p_{ij}-\alpha_{i}f_{j}-\beta_{i}g_{j})\dfrac{f_{j}}{\sigma_{ij}^{2}}=0 
		\end{equation}
		\begin{equation}
			\sum_{j}(p_{ij}-\alpha_{i}f_{j}-\beta_{i}g_{j})\dfrac{g_{j}}{\sigma_{ij}^{2}}=0 
		\end{equation}
		\begin{equation}
			\sum_{i}(p_{ij}-\alpha_{i}f_{j}-\beta_{i}g_{j})\dfrac{\alpha_{i}}{\sigma_{ij}^{2}}=0 
		\end{equation}
		\begin{equation}
			\sum_{i}(p_{ij}-\alpha_{i}f_{j}-\beta_{i}g_{j})\dfrac{\beta_{i}}{\sigma_{ij}^{2}}=0
		\end{equation}
		Equations above could be written in the form of matrices, and be used for iteration calculation of $f_{j}$, $g_{j}$, $\alpha_{i}$ and $\beta_{i}$:
		\begin{equation}
			\left(\begin{matrix}
				f_{j}' \\
				g_{j}' \\
			\end{matrix}\right)=
			\left(\begin{matrix}
				\sum_{i}\dfrac{\alpha_{i}^{2}}{\sigma_{ij}^{2}} & \sum_{i}\dfrac{\alpha_{i}\beta_{i}}{\sigma_{ij}^{2}}\\
				\sum_{i}\dfrac{\alpha_{i}\beta_{i}}{\sigma_{ij}^{2}} & \sum_{i}\dfrac{\beta_{i}^{2}}{\sigma_{ij}^{2}}\\
			\end{matrix}\right)^{-1}
			\left(\begin{matrix}
				\sum_{i} \dfrac{p_{ij} \alpha_{i}}{\sigma_{ij}^2} \\
				\sum_{i} \dfrac{p_{ij} \beta_{i}}{\sigma_{ij}^2} \\
			\end{matrix}\right)
			\label{equa:matrix1}
		\end{equation}
		
		\begin{equation}
			\left(\begin{matrix}
				\alpha_{i}' \\
				\beta_{i}' \\
			\end{matrix}\right)=
			\left(\begin{matrix}
				\sum_{j}\dfrac{f_{j}^{2}}{\sigma_{ij}^{2}} & \sum_{j}\dfrac{f_{j}g_{j}}{\sigma_{ij}^{2}}\\
				\sum_{j}\dfrac{f_{j}g_{j}}{\sigma_{ij}^{2}} & \sum_{j}\dfrac{g_{j}^{2}}{\sigma_{ij}^{2}}\\
			\end{matrix}\right)^{-1}
			\left(\begin{matrix}
				\sum_{j} \dfrac{p_{ij} f_{j}}{\sigma_{ij}^2} \\
				\sum_{j} \dfrac{p_{ij} g_{j}}{\sigma_{ij}^2} \\
			\end{matrix}\right)
			\label{equa:matrix2}
		\end{equation}
		
		The sub-integration $p_{ij}$ is an ``extended'' profile of $I$, $L$ and $V$: suppose the number of bins of profiles $I$, $L$ and $V$ is nbin, then the number of bins of $p_{ij}$ is $3\times$ nbin. The profiles of $I$, $L$ and $V$ are located at bin ranges of (0, nbin), (nbin, $2\times$ nbin) and ($2\times$ nbin, $3\times$ nbin) for $p_{ij}$.

		When deriving the $i$-th eigen profile, the $(i-1)$-th eigen profiles and mixture weights are set as initial conditions. Practically, before decomposing the eigen modes, we align the pulse profiles of $I$, $L$ and $V$ separately between different sub-integrations in the following way. The pulse profiles can be shifted from $t$ to $t+\Delta t$ using Fast Fourier Transformation (FFT):
		\begin{equation}
			\mathscr{F}^{-1}(\mathscr{F}(p)e^{i\omega \Delta t})=p(t+\Delta t)
			\label{equa:align}
		\end{equation}
  
		The time lag $\Delta t$ between two profiles $p_{1}$ and $p_{2}$ are derived below. The arguments of $\mathscr{F}(p_{1})$ and $\mathscr{F}(p_{2})$ could be written in the form:

        \begin{equation}
            \Phi_{1,k} = i\omega_{k} t
        \end{equation}
        \begin{equation}
            \Phi_{2,k} = i\omega_{k} (t+\Delta t)
        \end{equation}

        The $\omega_{k}$ is just 0, 1, 2, ... k ... as is defined in FFT algorithm. So through fitting $\Phi_{2,k} - \Phi_{1,k} = \omega_{k}\Delta t$, we can get $\Delta t$ and use it to align profiles by Equation~\ref{equa:align}.

\begin{figure}\centering
			\includegraphics[scale=0.26]{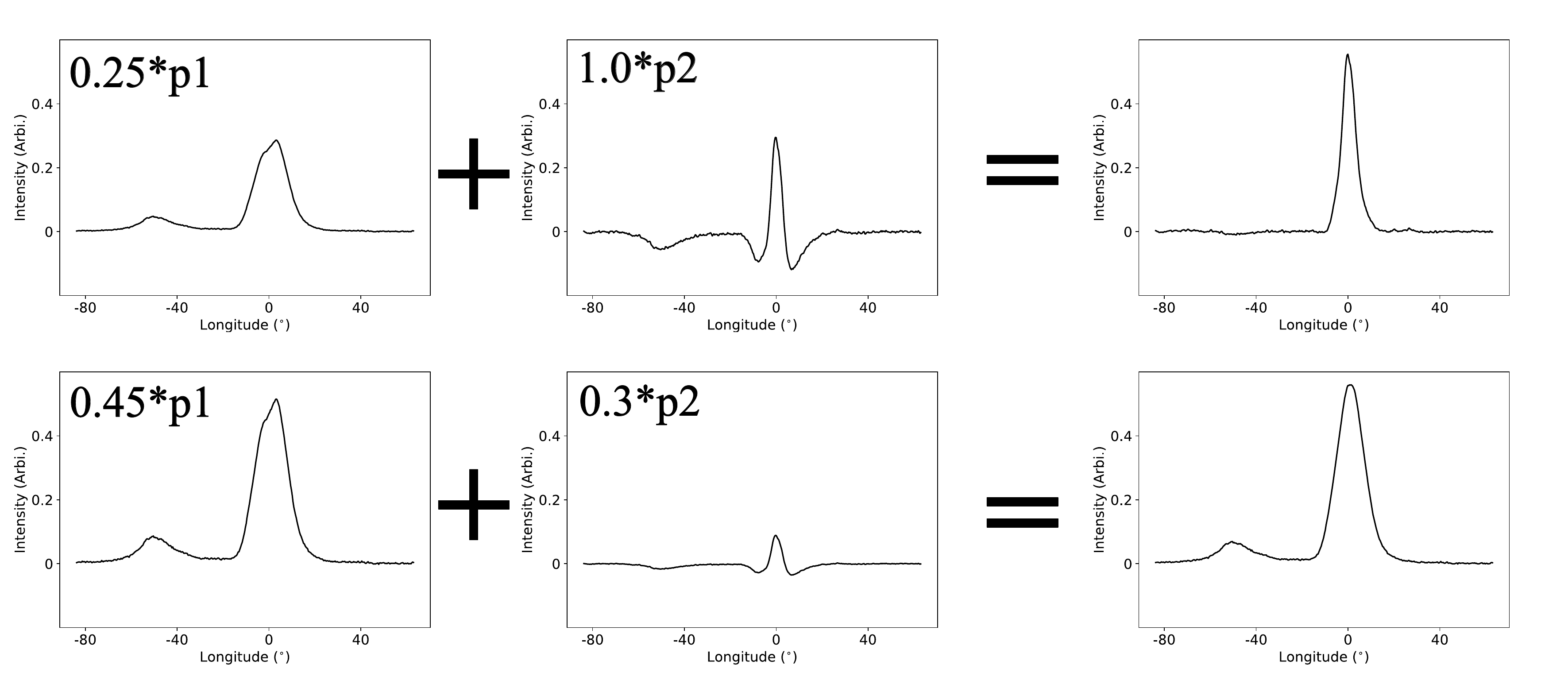}
			\caption{Two examples of mixing eigen mode profiles with different weights.}
            \label{fig:mixing_eg}
		\end{figure}

        For the pulse sequence of B0943$+$10, to improve signal-to-noise ratio, the decomposition is applied to all 4 epochs' sub-integrations made by 100 pulses every 5 pulses (e.g. 1-100, 6-105, ...) It must be emphasized that the eigen mode profiles decomposed from the pulse sequences are pure mathematical results. Though they do not have direct physical meaning, they together with the mixture weights do reveal some profile evolution properties appear in pulse sequences, as it will be shown in Section~\ref{sec:modeevo}. Two examples of mixing eigen mode profiles with different weights is presented in Figure~\ref{fig:mixing_eg}.

        \subsection{Mapping the radiation to the pulsar surface with emission geometry}\label{sec:geomap}

        Following is an introduction to the mapping method in \cite{Wang_2024}. The aim is to match the radiation phases to certain locations in the magnetosphere, given the inclination angle $\alpha$, the impact angle $\beta$, and a group of magnetic field lines (see Figure~\ref{fig:geomapping} for a schematic diagram). Here are some basic assumptions and approximations: (1) the magnetic field is pure dipole and centered at the pulsar center; (2) the radiation polarization mainly follows the adiabatic walking in \cite{1979ApJ...229..348C}, namely that the polarization main axes are either parallel (O mode) or perpendicular (X mode) to the magnetic field line plane (so the P.A. could be described with the RVM model); (3) the radiation frequency is considered simply as the critical frequency of curvature radiation.

        \par The first step of the algorithm is to choose appropriate groups of magnetic field lines, i.e. a group of magnetic azimuths and magnetic colatitudes ($\Phi_{i}$,$\Theta_{i}$) ($i=1,2,3,...$, spherical coordinates regarding the magnetic axis 's footprint as ($0$,$0$)) of the magnetic fields' footprints on the polar cap for calculation. Here the concepts of the inner annular gap (IAG) and the inner core gap (ICG), which are introduced in \cite{2004ApJ...606L..49Q} to make the magnetospheric current closed, are adopted. In the polar cap region, IAG means the region inside the feet of critical magnetic field lines, and ICG means the region between the feet of critical field lines and the feet of last open field lines. The region too close to magnetic axis has too straight magnetic field line to produce significant radiation \citep[e.g.,][]{2018PhyU...61..353B}, and the region too close to the last open field lines near the pulsar surface may have not large enough parallel electric field for particle acceleration (based on RS model \citep{1975ApJ...196...51R}). So the representative field lines we choose are not too close to the above two regions. Given the inclination angle $\alpha$ and the pulsar rotational period, the feet locations of critical field lines and last open fields could be calculated, which are just the red line and the black line in Figure~\ref{fig:geomapping} (ii). The ICG's representative field lines' feet are chosen to have 2/3 distances of the critical field lines' feet to the magnetic axis, and the IAG's representative field lines' feet are chosen to have equal distances to the critical field lines' feet and to the last open field lines' feet.

        \par After the representative groups of magnetic field lines are determined, the second step is to match the radiation at certain rotation phases (or longitude, $\phi_{i}$) to corresponding magnetic field lines ($\Phi_{i}$,$\Theta_{i}$), and to find the emission point. The radiating particles' Lorentz factors $\gamma$ are usually about $10^{2}\sim 10^{3}$, so the emission cones are small ($1/\gamma\ll 1$), and thus we simply regard the tangential direction of the magnetic field line at the emission point as the emission direction. The second step is equivalent to find a magnetic field line with footprint at ($\Phi_{i}$,$\Theta_{i}$) that has a tangential direction (at the emission point) same as the direction of the line of sight (LOS), at longitude $\phi_{i}$. At the same time, the emission point is determined, so the emission height, i.e. the radial distance from emission point to the pulsar central point, could be directly calculated.

        \par The third step is estimating the Lorentz factor of radiating particles at emission points. At emission points, the curvature radii $\rho_{c}$ of dipole magnetic field lines are known. Then follow the third assumption at the beginning of this section, we equal curvature radiation's critical frequency to the radiation frequency $\nu$ (in FAST's case, we use 1250~MHz), and thus the Lorentz factors are just:
        \begin{equation}
			\gamma=\left(2\pi\nu\cdot \dfrac{2\rho_{c}}{3c} \right)^{1/3}
			\label{eq:freq_c}
		\end{equation}

		\section{Results}\label{sec:sec3}
\begin{figure}\centering
			\includegraphics[scale=0.36]{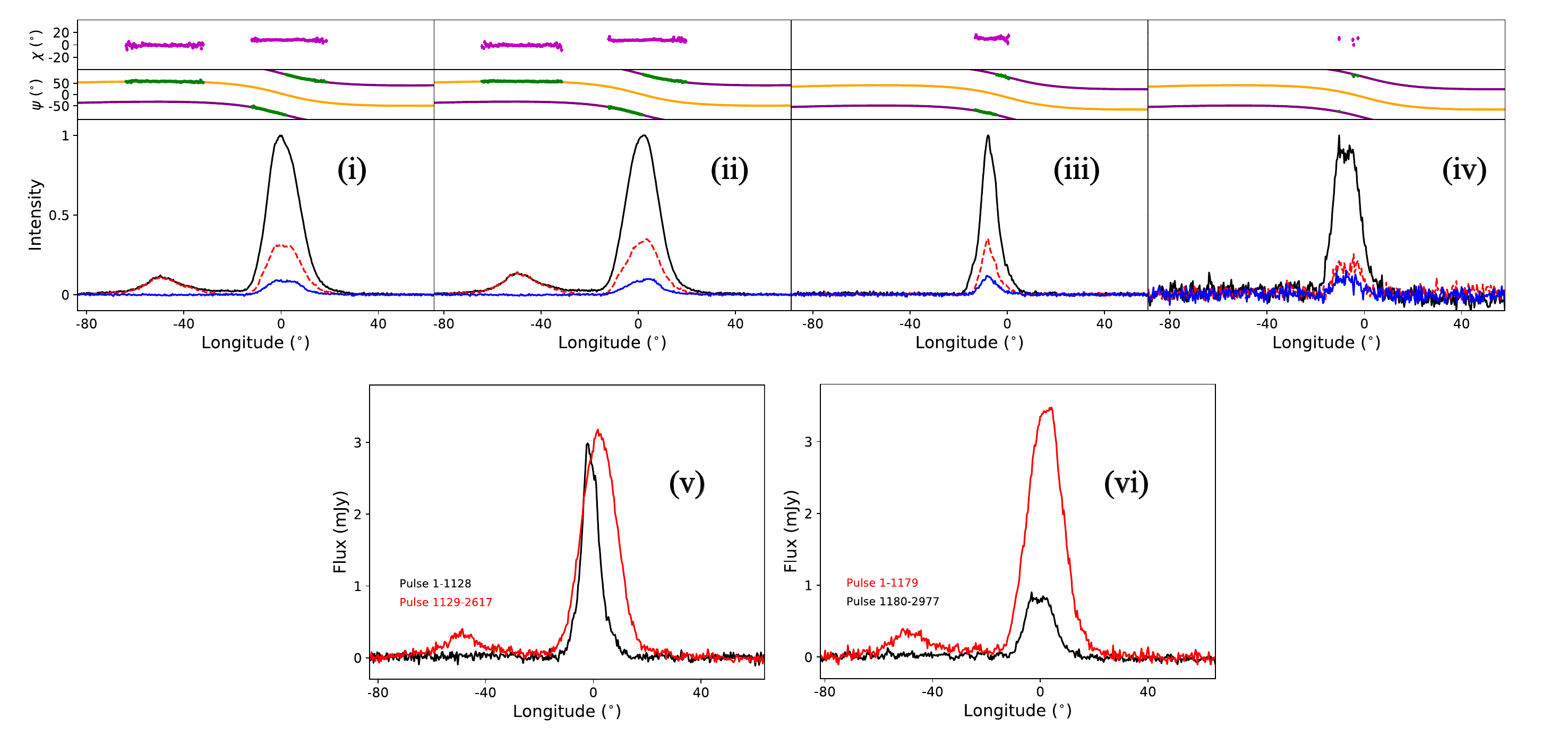}
			\caption{Upper: integrated profiles of (i) all 14521 pulses in total 4 epochs; (ii) all 9237 Q mode pulses; (iii) all 3486 B mode pulses; (iv) all 1798 ``B' mode'' pulses. Black line - intensity ($I$); red dash line - linear polarization intensity ($L=\sqrt{Q^{2}+U^{2}}$); blue line - circular polarization intensity ($V$). $I$, $L$ and $V$ are normalized by the maximum intensity of the respective profiles. Green dots in $\psi$ panel with errorbars - polarization position angles (P.A., $\psi=0.5\arctan(U/Q)$); Orange and purple lines in $\psi$ panel - RVM curves with 90$^{\circ}$ displacement, of which the details are given in Section~\ref{sec:rvm}. Magenta dots in $\chi$ panel with errorbars - elliptical angel (E.A., $\chi=0.5\arcsin(V/{\sqrt{U^{2}+Q^{2}+V^{2}}})$). Only P.A. and E.A. at longitudes where $L/\sigma_{L}>10$ are shown in the plots. Lower: comparison of profiles' flux density before and after mode switches in epoch (b) (plot (v)) and in epoch (c) (plot (vi)). The flux density is estimated with Equation~\ref{eq:flux}.}
            \label{fig:int_prof}
		\end{figure}

		The integrated pulse profiles of all 4 epochs' pulses, all Q mode pulses, all B mode pulses and the ``B' mode'' of PSR~B0943$+$10 are shown in Figure~\ref{fig:int_prof}, where the polarization position angels (P.A.) and the ellipticity angles (E.A.) of on-pulse regions are also presented. Equation~\ref{eq:PA_def} and~\ref{eq:EA_def} are the definitions of P.A. and E.A.~\citep[e.g.,][]{1979rpa..book.....R}. P.A. together with E.A. curves are equivalent to the profile's polarization evolution track on the Poincare sphere, which could provide a more complete view of the polarization's evolution with phase~\citep[e.g.,][]{2004A&A...421..681E,2020MNRAS.495L.118D}.
        \begin{equation}
            \text{P.A.} = \psi= \dfrac{1}{2}\arctan\dfrac{U}{Q}
            \label{eq:PA_def}
        \end{equation}
        \begin{equation}
            \text{E.A.} =\chi= \dfrac{1}{2}\arcsin\dfrac{V}{P}=\dfrac{1}{2}\arcsin\dfrac{V}{\sqrt{Q^{2}+U^{2}+V^{2}}}
            \label{eq:EA_def}
        \end{equation}
        The profiles' P.A. curves could be well fitted with two orthogonal RVM curves, of which the details will be presented in Section~\ref{sec:rvm}. All pulse profiles' longitude $0^{\circ}$ is chosen where the fitting RVM curves are steepest, i.e. the $\phi = \phi_{0}$ point of Equation~\ref{eq:RVM}. Following is the radiation modes identification and analysis of the 4 epochs' profile evolution.

		\begin{figure}\centering
			\includegraphics[scale=0.2]{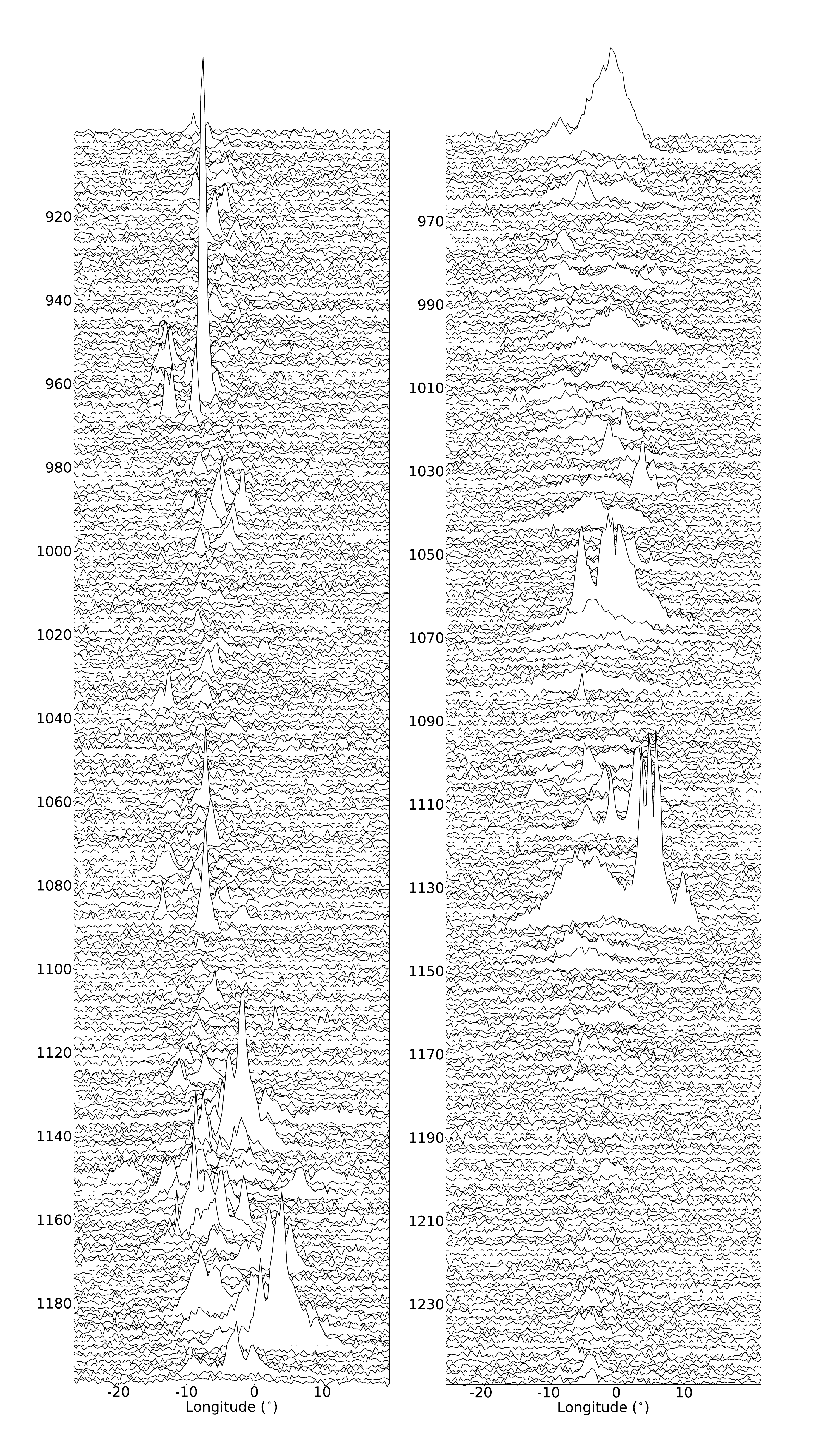}
			\caption{Waterfall plots of pulses around mode switches in epoch (b) (left) and in epoch (c) (right). Horizontal axis: phase (0-1 for a period); vertical axis: number of pulses ($N$, top to bottom). The intensity is normalized with the off-pulse noise, for each pulse.}\label{fig:waterfall_2}
		\end{figure}

		\subsection{Modes and profiles' evolution}\label{sec:modeevo}

\begin{figure}
			\centering
			\includegraphics[scale=0.2]{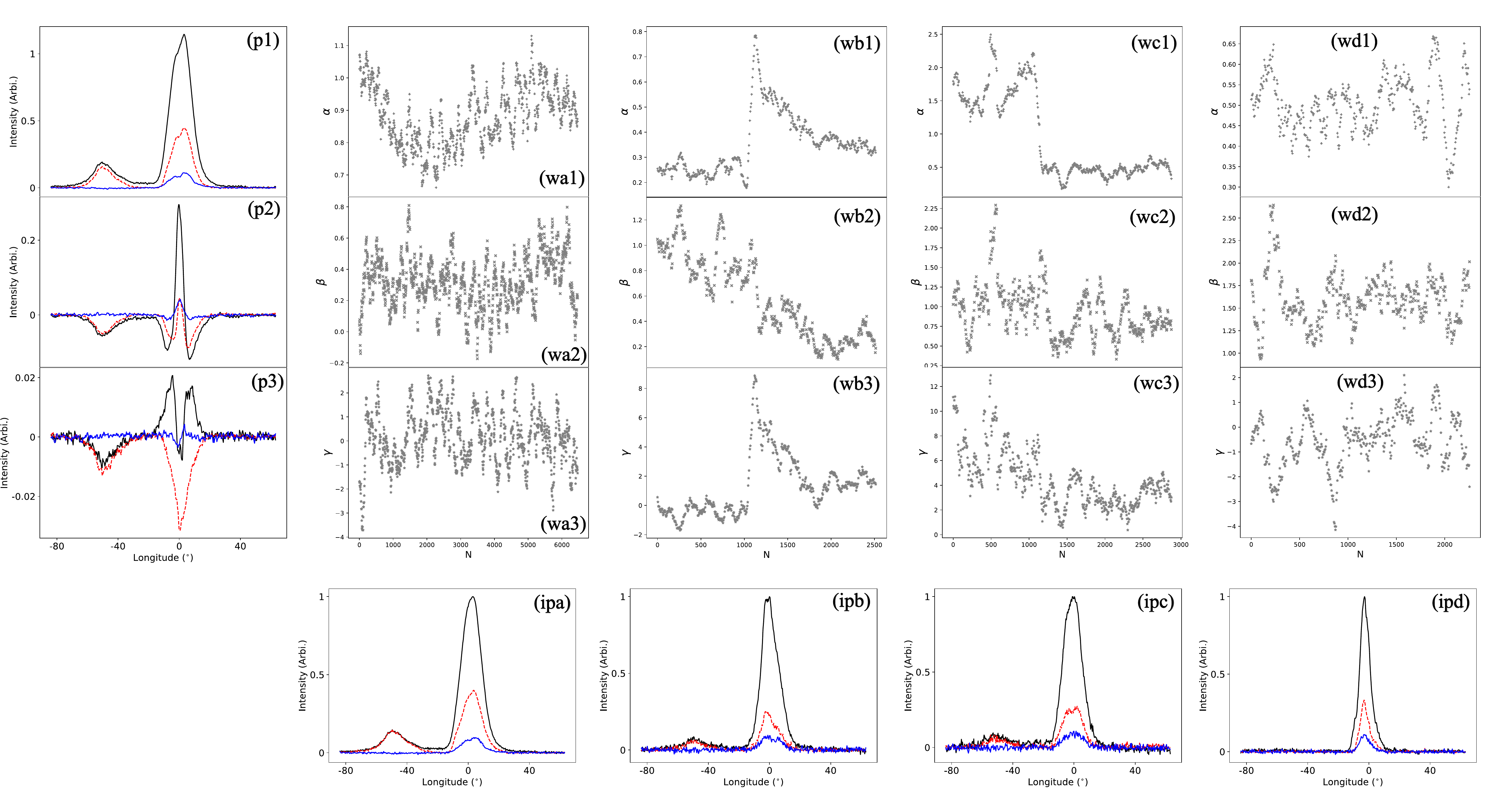}
		    \caption{(p1), (p2) and (p3): three decomposed eigen profiles. (ipa), (ipb), (ipc) and (ipd): the intensity ($I$), linear polarization ($L$) and circular polarization ($V$) of epoch (a)(b)(c)(d)'s integrated profiles. Lines' meanings are same as Figure~\ref{fig:int_prof}. (wa1) ... (wd3): corresponding mixture weights of (p1), (p2) and (p3) versus pulse numbers, of (a)(b)(c)(d) 4 epochs. }\label{fig:decom_result_1}
		\end{figure}

        We begin the decomposition algorithm from $i=1$, namely only 1 eigen mode ($i=1$), by setting $f_{0j}=0$ and setting random initial values of the mixture weight. With the first eigen mode derived, the calculation of two eigen modes begins by setting the first eigen profiles and weights as initial conditions. For deriving three eigen modes the process is same. After all 3 profiles are derived, we make them mutually orthogonal through Gram-Schmidt orthogonalization, and calculate their corresponding mixture weights. The results, as well as all 4 epochs' integrated profiles, are shown in Figure~\ref{fig:decom_result_1}. The first eigen profile (p1) has a wide main pulse component and a positive precursor component, and the second profile (p2) has a narrow main pulse component and a negative precursor component. The superposition of the first two eigen modes is shown as an example in Figure~\ref{fig:mixing_eg}. The third eigen profile (p3) shows an anti-correlation between the precursor's and the main pulse's intensity, which has been observed in B0943$+$10's Q mode pulses~\citep{2010MNRAS.404...30B}.

\begin{figure}
			\centering
			\includegraphics[scale=0.27]{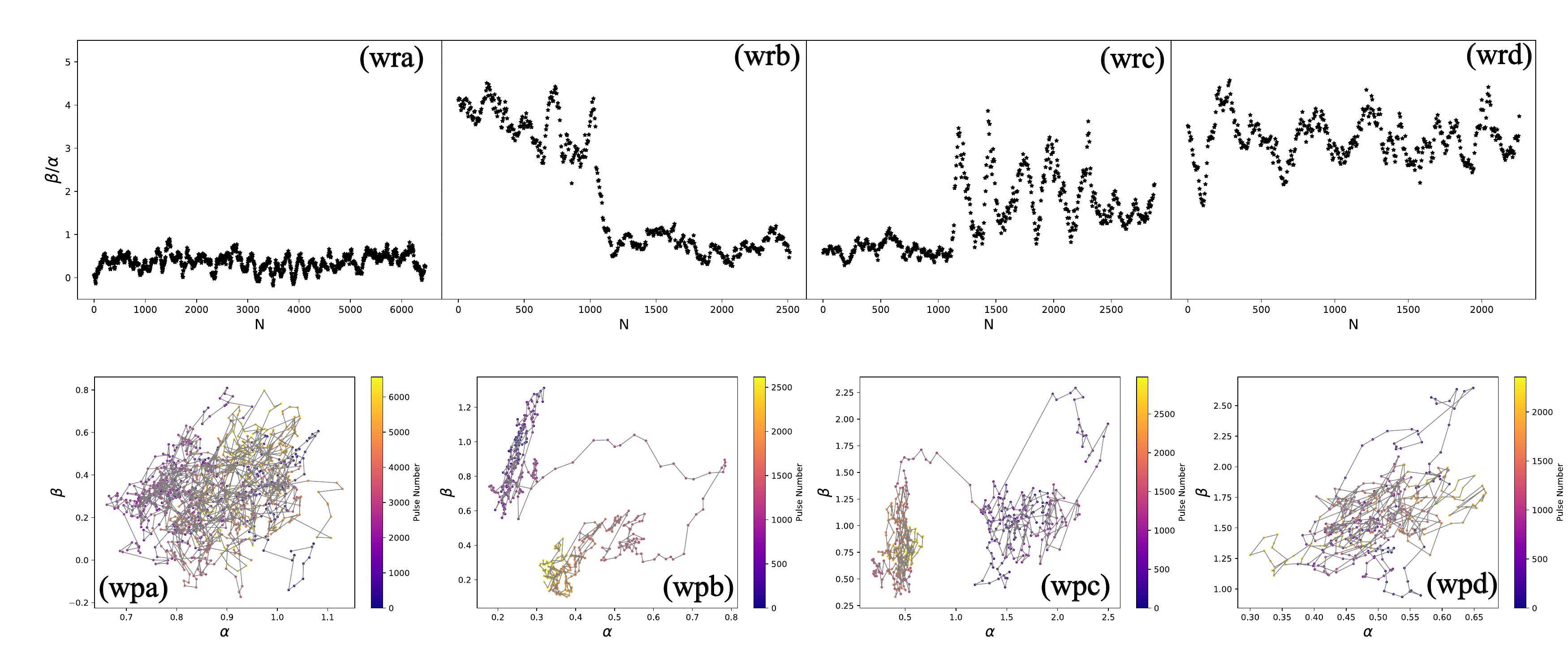}
        \caption{(wra), (wrb), (wrc) and (wrd): the ratios of the mixture weights of (p2) and (p1) (in Figure~\ref{fig:decom_result_1}) versus pulse numbers. (wpa), (wpb), (wpc) and (wpd): mixture weights of (p1) and (p2) shown in 2D plane ("phase space"). The colorbar is used to mark the pulses' evolution on this $\alpha$ - $\beta$ diagram with the pulse number's increase. The pulse sequence begins at deep blue dots and ends at bright yellow dots. }\label{fig:decom_wr}
		\end{figure}

        For a better description of the mode profile evolution process, especially the mode switch, some extra plots on the ratios of the first two eigen modes' weights ($\beta/\alpha$) and the pulse sequences' evolution on a $\alpha$ - $\beta$ plane (2D phase space) are drawn and shown in Figure~\ref{fig:decom_wr}. The $\beta/\alpha$s of one epoch could be compared with the other epochs'. Following is the analysis to all 4 epochs' profile evolution based on Figure~\ref{fig:decom_result_1} and Figure~\ref{fig:decom_wr}.

        \begin{figure}\centering
			\includegraphics[scale=0.18]{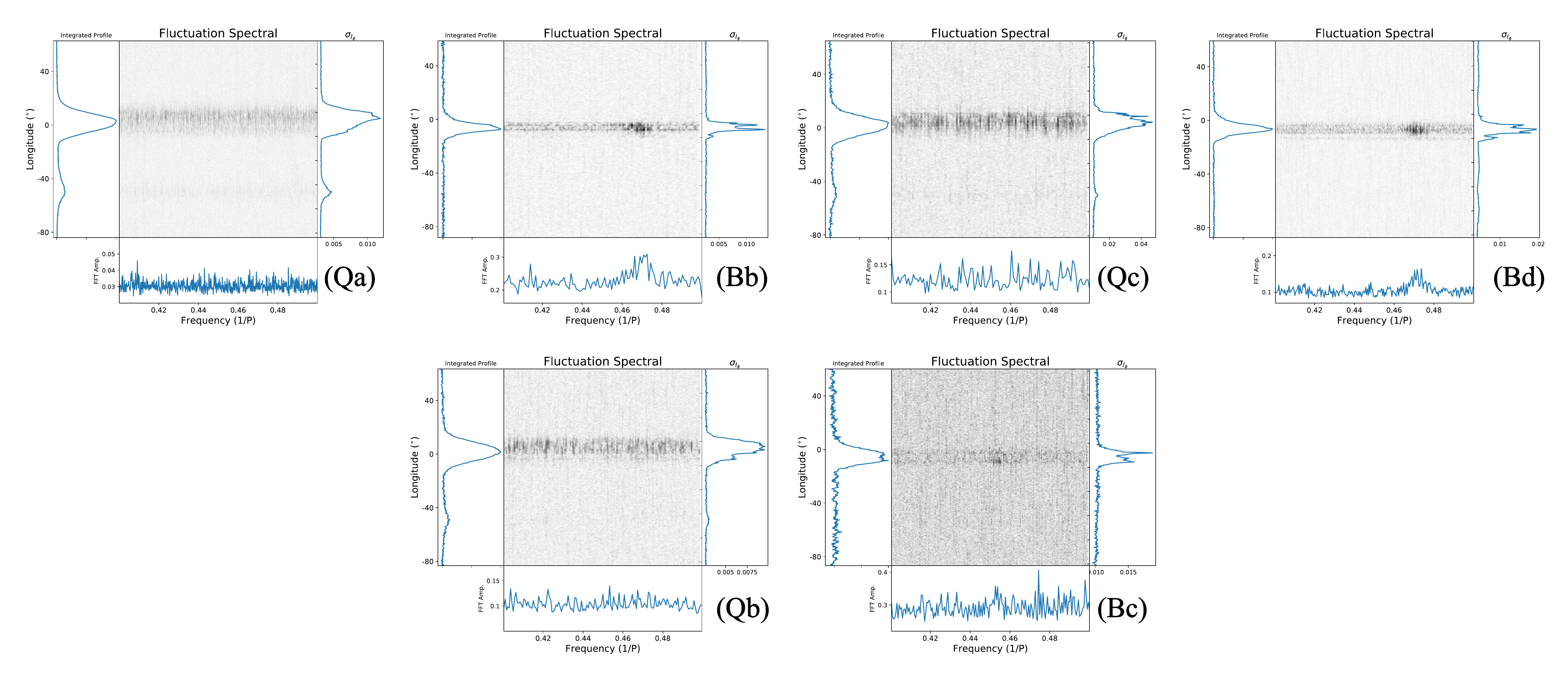}
			\caption{LRFS of (Qa) 6569 pulses in epoch (a); (Bb) 1128 B mode pulses in epoch (b); (Qb) 1489 Q mode pulses in epoch (b); (Qc) 1179 Q mode pulses in epoch (c); (Bc) 1798 ``B' mode'' pulses in epoch (c); (Bd) 2358 B mode pulses in epoch (d). For each LRFS plot, the body show FFT amplitude of each phase and frequency; the left-hand panel shows the integrated profile of the observed pulses; the right-hand panel shows the standard deviation of single pulse intensity on every phase bins; the lower panel shows the integral spectra.\label{fig:LRFSs}}
		\end{figure} 
  
		\subsubsection{Epoch (a): pure Q mode}\label{sec:220517}
		The epoch (a)'s integrated profile (ipa) in Figure~\ref{fig:decom_result_1} has a highly linearly polarized ($\sim 100$\%) precursor component at about 0.14 phase earlier than the main component. This feature is consistent with B0943$+$10's Q mode profile reported by other observations \citep[e.g.,][]{2010MNRAS.404...30B,2013Sci...339..436H}. The mixture weights (wa1), (wa2) and (wa3) in Figure~\ref{fig:decom_result_1} and the first two weights' ratio $\beta/\alpha$ (wra) in Figure~\ref{fig:decom_wr} show no sudden jumps, and the pulse sequence's evolution in $\alpha$ - $\beta$ plane ((wpa) in Figure~\ref{fig:decom_wr}) follows only one single patch. Besides, the longitude-resolved fluctuation spectra (LRFS) of totally 6569 pulses in epoch (a) is given in Figure~\ref{fig:LRFSs}. There's no periodic fluctuation in the pulse components, which is typical for B mode pulses~\citep[e.g.,][]{deshpande2001}. All these facts indicating that pulses in epoch (a) data are pure Q mode.

		\subsubsection{Epoch (d): pure B mode}\label{sec:230827}
		The epoch (d) profile (ipd) in Figure~\ref{fig:decom_result_1} has only one pulse component, and it is manifestly narrower than the Q mode pulse profile of epoch (a). The mixture weights (wb1), (wb2) and (wb3) in Figure~\ref{fig:decom_result_1} and the first weights' ratio $\beta/\alpha$ ((wrb) in Figure~\ref{fig:decom_wr}) show no sudden jumps. And there's one patch on the $\alpha$ - $\beta$ plane ((wpb) in Figure~\ref{fig:decom_wr}). Figure~\ref{fig:LRFSs} shows epoch (d) pulses' LRFS (Bd), where there is an FFT amplitude peak at about 0.47 cycle/P, only in the pulse phase range. This fluctuation frequency is consistent with results of former studies on B0943$+$10's B mode~\citep[e.g.,][]{deshpande2001}. All these facts indicate that pulses in epoch (d) data are pure B mode.

        It's worth noticing that in Figure~\ref{fig:LRFSs} at longitude $\phi \approx -15^{\circ}$, a distinct line appears in fluctuation spectra, and there is a peak of standard deviation of single pulses' intensity. Such phenomena indicates an extra component in B mode pulses. From previous observations we learn that the B mode profile is two-hump like under very low frequency frequency ranges~\citep[e.g.,][]{2014A&A...572A..52B, 2018A&A...616A.119B}. When frequency increases, the B mode profile changes into one-hump. So the B mode extra component shown in fluctuation spectra (as well as in single pulses' P.A., E.A. and $L/I$ distributions shown in Figure~\ref{fig:PALI}) might just correspond to one of the two components under low frequencies.

    		\subsubsection{Epoch (b): B-to-Q switch}\label{sec:220902}
		Figure~\ref{fig:waterfall_2} has clearly shown a mode switch from narrow pulses to wide pulses in epoch (b). The mixture weights (wb1), (wb2) and (wb3) in Figure~\ref{fig:decom_result_1} all show sudden jumps around $N\sim 1100$, where the first two weights' ratio $\beta/\alpha$ ((wrb) in Figure~\ref{fig:decom_wr}) switches from larger values to smaller values, indicating a B-to-Q mode switch. Besides, it's worth noting that the evolution pattern on $\alpha$ - $\beta$ plane (wpb) in Figure~\ref{fig:decom_result_1} looks like a strange attractor of a chaotic system.

		LRFS of the two modes' single pulses are shown in Figure~\ref{fig:LRFSs}. The first mode's FFT amplitude has a peak at about 0.47 cycle/P while the second dosen't have. Comparing to results of epoch (a) (Section~\ref{sec:220517}) and epoch (d) (Section~\ref{sec:230827}), the conclusion is that the first mode is B mode and the second mode is Q mode. The extra component at $\phi \approx -15^{\circ}$ also appears in Figure~\ref{fig:LRFSs}.
		
\begin{figure}\centering
			\includegraphics[scale=0.49]{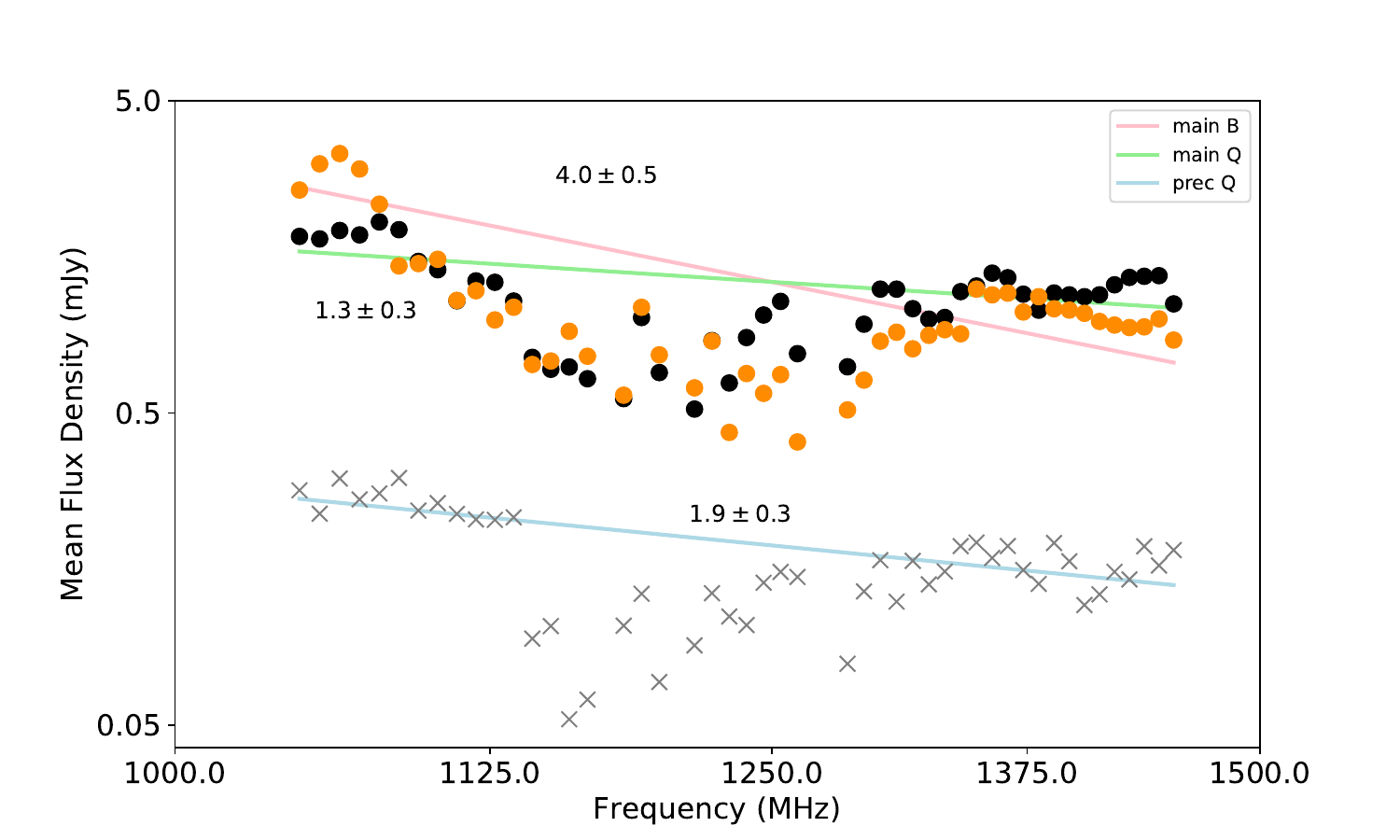}
			\caption{Spectra of B mode's main pulse component (orange dots), Q mode's main pulse component (black dots) and Q mode's precursor component (grey crosses) in epoch (b). Each point represents the mean flux density of one frequency channel (64 channels in total), and is averaged within the component's on-pulse phase range. All the dots and crosses are actually with errorbars. The spectra are fitted with power law functions ($C\nu^{-\alpha}$), which are lines (pink, light green and light blue) under logarithmic coordinates. The best fitted spectral indices ($\alpha$s) are marked near the lines. All three spectra dip in the middle range, which is probably caused by scintillations.}\label{fig:spectral}
		\end{figure}

		Knowing FAST's system temperature ($T_{\text{sys}}$) and gain ($G$) (using data from \cite{2020RAA....20...64J}), the flux density of the two modes could be estimated:
		\begin{equation}
			I_{\phi} = \dfrac{\text{SNR}\cdot T_{\text{sys}}}{G}\sqrt{\dfrac{\text{nbin}}{2\times \text{bandwidth} \times t_{\text{obs}}}}
        \label{eq:flux}
		\end{equation}
		The comparison of the two modes' flux density is shown in Figure~\ref{fig:int_prof}. While Q mode being apparently dimmer under low frequencies (like Figure 1 in~\cite{2011MNRAS.418.1736B}), under 1-1.5GHz Q mode is almost as bright as B mode, indicating different frequency evolutions of the two modes. The spectra of the main pulse components of B and Q mode, and of the precursor component of Q mode, is shown in Figure~\ref{fig:spectral}. B mode has a larger spectral index, while for Q mode the precursor's and the main pulse's spectral indices have no significant differences.

		\subsubsection{Epoch (c): Q-to-``B' mode'' switch}\label{sec:230816}
		Figure~\ref{fig:waterfall_2} (right) shows the waterfall plots around mode switch at $N\approx 1180$ for epoch (c). Sudden jumps also happens around this pulse number in the mixture weights' evolution (wc1), (wc2) and (wc3), and two distinct patches also appear in $\alpha$ - $\beta$ plane ((wpc) in Figure~\ref{fig:decom_wr}). The first two weights' ratio $\beta/\alpha$ ((wrc) in Figure~\ref{fig:decom_wr}) is generally small for the first half of pulses, but then oscillates and slowly moves to larger values after $N\sim 1000$.

		LRFS of the two modes are shown in Figure~\ref{fig:LRFSs} (Qc) and (Bc). Comparing with the results of epoch (a), epoch (b) and epoch (d), we can conclude that the first mode in epoch (c) is Q mode. The second mode, whose integrated profile is shown in Figure~\ref{fig:int_prof} as plot (iv), has a narrow main pulse component, and has no precursor. There is possible fluctuation at about 0.45 cycle/P, but not very clear. The profile after mode switch is significantly different from B mode's profile in Figure~\ref{fig:int_prof}. We mark the second mode as ``B' mode'' for the present.
		
		The flux density of the two modes on epoch (c) is also estimated in the same way of Section~\ref{sec:220902}, and the result is shown in Figure~\ref{fig:int_prof}. Both from the flux density comparison and the waterfall plot, ``B' mode'''s intensity is much weaker than Q mode's.
		
		\subsection{Radiative geometry and polar cap mapping}\label{sec:PA_geo}
  
        \subsubsection{RVM fitting of the P.A. curve}
        \label{sec:rvm}
\begin{figure}
			\centering
			\includegraphics[scale=0.4]{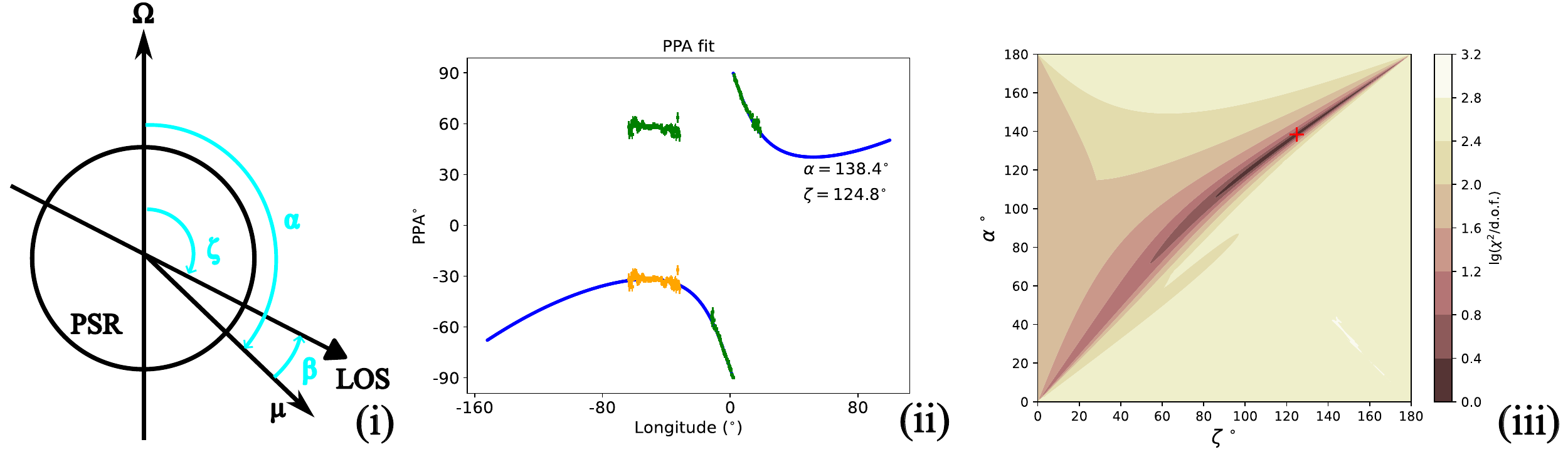}
			\caption{Plots on radiation geometry fitting. Plot (i): sketch of relations of $\alpha$, $\beta$ and $\zeta$ in the fitting described in Section~\ref{sec:rvm}. Plot (ii): RVM fitting of the modified P.A. curve (shift the precursor P.A. dots 90$^{\circ}$ downward (downward = upward) (colored orange)). The original P.A. curve of the total integrated profile in Figure~\ref{fig:int_prof} (i) are presented by the green dots with errorbars. Plot (iii): the common logarithm values of chi-square (divided by degree of freedom) of fitting RVM curve of given ($\alpha$,$\zeta$) to the modified P.A. curve. The red ``+'' sign marks the ($\alpha$,$\zeta$) where we get the smallest $\chi^{2}$.\label{fig:geofit}}
		\end{figure}

        In Figure~\ref{fig:int_prof}, the P.A. curves show ``S'' shape in the main pulse longitude range, but are almost flat in the precursor pulse longitude range. RVM model (in the form of Equation~\ref{eq:RVM}) could well describe the ``modified'' P.A. curves, after shifting the P.A.s of the precursor component by an angle of 90$^{\circ}$ (shown in Figure~\ref{fig:geofit} (ii)). The wide longitude range of on-pulse region makes RVM fitting possible. When making least square fitting with Equation~\ref{eq:RVM}, $\alpha$ is assumed to be in the range ($0^{\circ}$, $180^{\circ}$). All four epochs' pulse profiles could be fitted with the same $\alpha$, $\beta$ with different $\psi_{0}$s (Figure~\ref{fig:int_prof}). The fitting result is that the inclination angle $\alpha=138\pm 2(^{\circ})$ and the impact angle $\beta=-14\pm 4(^{\circ})$, i.e. the LOS is equatorward. For making comparison with results of other literature, the pair of values $\alpha'=42\pm 2(^{\circ})$ and $\beta'=14\pm 4(^{\circ})$ should also be considered, according to~\cite{2001ApJ...553..341E}. The two angles could be used to calculate the maximum gradient of the P.A. curve $R_{\text{P.A.}}$, which is in the form of $-|\sin\alpha/\sin\beta|$. And the value is $R_{\text{P.A.}}=-2.76^{\circ}/^{\circ}$, which is in accord with values around $-2.4^{\circ}/^{\circ} \sim -3.6^{\circ}/^{\circ}$ given by \cite{1998JApA...19....1S,deshpande2001,2010MNRAS.404...30B} (also see Table~\ref{tab:R_PA} for a comparison between our work and former papers, in observation frequencies, $R_{\text{P.A.}}$, $\alpha$ and $\beta$). The above fitting results are also checked by the chi-square calculation result shown in Figure~\ref{fig:geofit}: for each pair of ($\alpha$,$\zeta$), we calculate the least chi-square value when changing ($\psi_{0}$, $\phi_{0}$) ($\phi_{0}$ is limited between $-90^{\circ}$ and $90^{\circ}$). The RVM curve does fit the P.A.s well within a small ($\alpha$,$\zeta$) range around ($138^{\circ}$, $125^{\circ}$). P.A.s of the main pulse and the precursor are fitted with two RVM curves displaced by $90^\circ$ respectively, indicating that the main pulse and the precursor are orthogonal in linear polarization.
		\begin{equation}
			\tan(\psi-\psi_{0})=\dfrac{\sin(\phi-\phi_{0})\sin\alpha}{\sin\zeta \cos\alpha-\cos\zeta \sin\alpha\cos(\phi-\phi_{0})}
			\label{eq:RVM}
		\end{equation}
        To check the geometry by fitting the spectral evolution of pulse width, we calculate the profile width of epoch (a) and (d) under divided frequency bands, but it shows that the width's spectral evolution is not monotonic.

        \begin{table}
			\centering
			\renewcommand\arraystretch{1.2}
			\caption{A comparison between different papers' results on maximum P.A. gradient, inclination angle $\alpha$ and impact angle $\beta$. The corresponding observation frequencies are also presented.\label{tab:R_PA}}
			
			\begin{tabular}{lcccc} 
				\hline
				\hline
				SOURCE: & \cite{1998JApA...19....1S} & \cite{deshpande2001} & \cite{2010MNRAS.404...30B} & This paper \\
                \hline
                FREQUENCY (MHz): & $102.5$ \& $430$ & $102.5$ \& $430$ & $327$ \& $430$ & $1250$ \\
				\hline
                $R_{\text{P.A.}}( ^{\circ}/^{\circ})$: & $-2.4 \sim -3.6$ & $-2.7$ & $-3.0$ & $-2.76$ \\
				\hline
                $\alpha (^{\circ})$:&-& 11.58  &- & 42 (138) \\
                \hline
                $\beta (^{\circ})$:&-& -4.29 &-& 14 (-14)\\
                \hline
			\end{tabular}
		\end{table}
  
        \subsubsection{Radiation mapping onto the polar cap}
        \label{sec:result_map}
\begin{figure}\centering
            \includegraphics[scale=0.38]{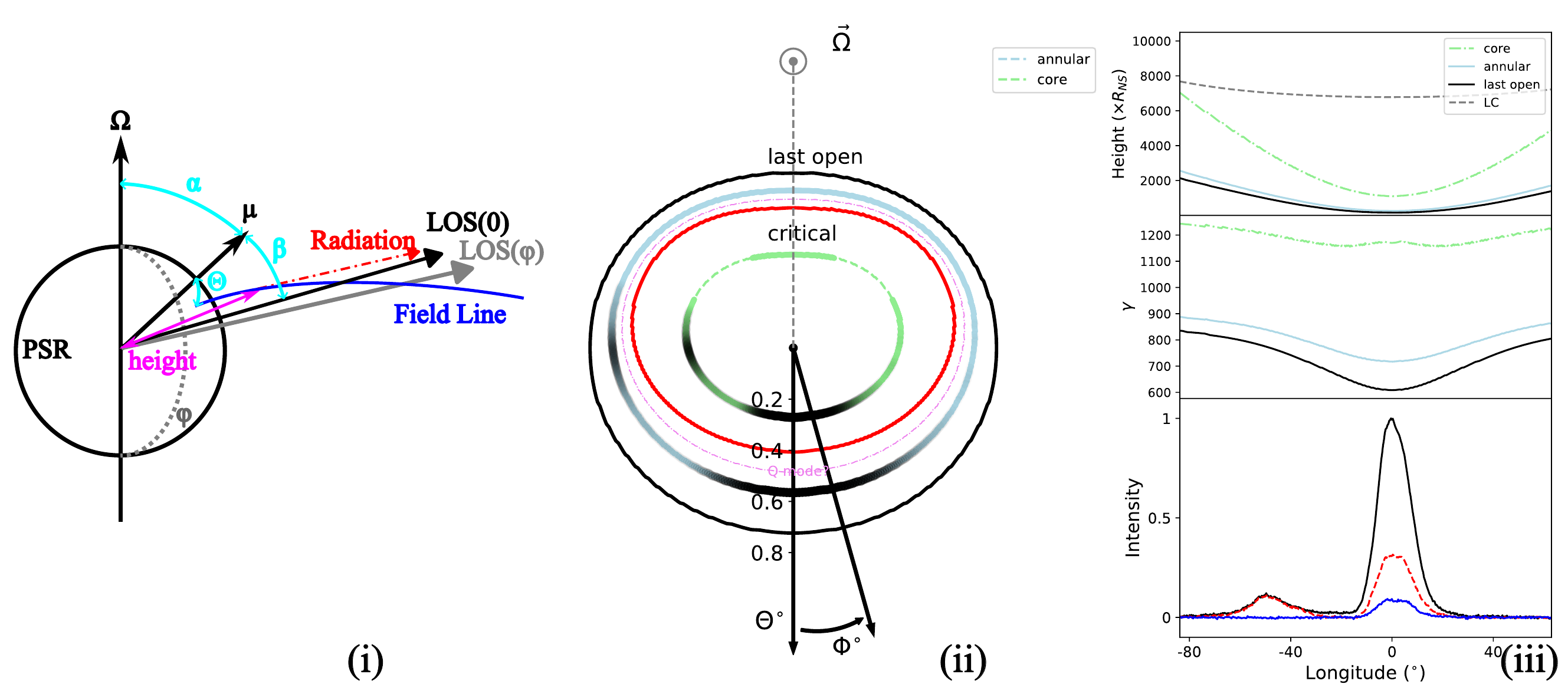}
            \caption{(i) a sketch for describing the mapping process in Section~\ref{sec:geomap}; (ii): B0943$+$10's radiation mapped to the surface polar cap region. The magnetic axis is right in the middle. The numbers 0.2, 0.4, 0.6, 0.8 represent the magnetic colatitudes $\Theta(^{\circ})$, and $\Phi$ means the magnetic azimuths. Black solid line: feet of last open field lines (margin of the polar cap); red solid line: feet of critical field lines (margin of the core gap region); light blue line: feet of representative magnetic field lines in annular gap region, having equal distances to last open field lines' feet and critical field lines' feet; green line: feet of representative magnetic field lines in core gap region, having $2/3$ distances of critical field lines' feet to the magnetic axis. The dashed line's part is invisible to our line of sight because of the radiating height exceeding the light cylinder; light purple dash-dotted line: possible origins of Q mode radiation particles (will be discussed in Section~\ref{sec:discussion_geo} and~\ref{sec:opms_discussion}), having equal distances to the light blue line and the solid red line; thick black lines with opacity: radiation mapped to those typical field lines' feet. The opacity is positively correlated to the total intensity observed by FAST in the integrated profile shown in Figure~\ref{fig:int_prof}. (iii) Estimation of emission heights and radiating particles' Lorentz factors. Upper panel: radiation height, the distance from the radiation position to the pulsar's central point. Middle panel: Lorentz factors of radiation particles. Light green dotted dash line: heights and Lorentz factors calculated with the representative field lines in core region (the light green line in (ii))); Light blue line: heights and Lorentz factors calculated with the representative field lines in annular region (the light blue line in (ii))); Black line in upper panel and middle panels: heights and Lorentz factors calculated with the last open field lines; Gray dashed line: distances from the pulsar's central point to light cylinder in the radiating directions at certain phases. Lower panel: the profile in Figure~\ref{fig:int_prof}'s (1).\label{fig:geomapping}}
		\end{figure}

        Methods introduced in Section~\ref{sec:geomap} are applied to map radiation to pulsar surface geometrically, with the help of two representative groups of magnetic field lines, and the result is shown in Figure~\ref{fig:geomapping}. B0943$+$10 has a small polar cap ($r < 1^{\circ}$), and the spark discharge point of precursor component radiation is generally located further away from the magnetic axis than that of main pulse radiation. 
		
		Based on the mapping result, the radiation height (namely the distance from the tangential point of line-of-sight and magnetic field line of certain phase to the center of the pulsar) and radiation particles' Lorentz factors ($\gamma$) (based on curvature radiation's critical frequency equation, see Equation~\ref{eq:freq_c}) could also be estimated, and the results are shown in Figure~\ref{fig:geomapping}. For B0943$+$10's case, if the radiating particles originate from the annular region, they tend to radiate at a lower height with smaller Lorentz factor. And if both the main pulse component and the precursor component are radiated by particles from the same region, the precursor component is emitted higher by particles with larger Lorentz factor.
		
		\subsection{Single pulses' polarization}\label{sec:single_pol}
  
\begin{figure}\centering
			\includegraphics[scale=0.23]{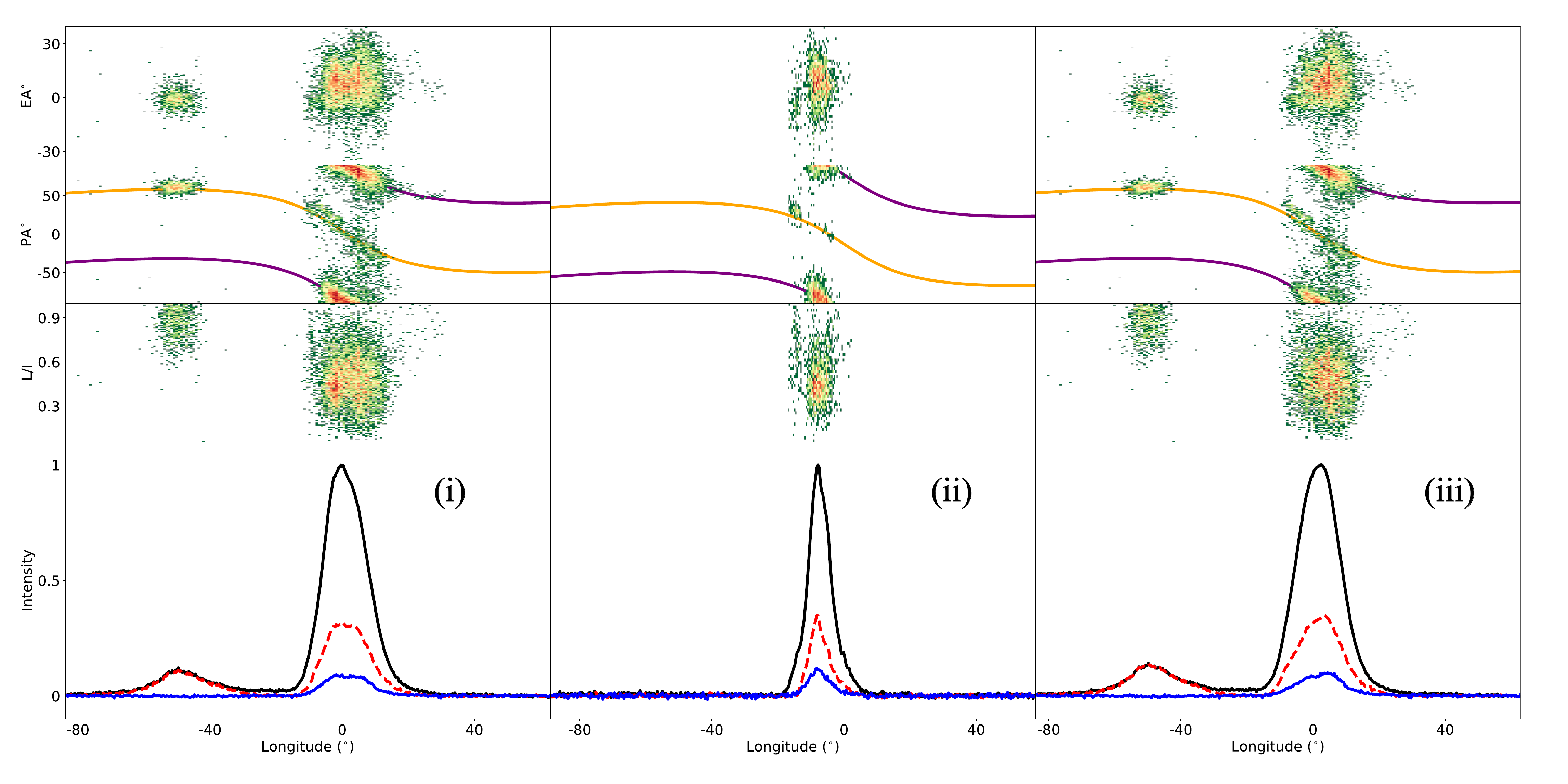}
            \caption{Distributions of P.A., E.A. and $L/I$ of all 4 observation epochs' 14519 pulses (i), all 9237 Q mode pulses (ii) and all 3486 B mode pulses (iii). Panel 4 (1-4 from up to bottom): the integrated profiles of all pulses taken into account, where lines' meanings are same as Figure~\ref{fig:int_prof}; Panel 3: the distributions of $L/I$; Panel 2: the distributions of P.A.. Panel 1: the distributions of E.A.. Definitions of P.A. and E.A. follow Equation~\ref{eq:PA_def} and~\ref{eq:EA_def}. For panel 1, 2 and 3, the distribution is more concentrated when color changes from green, yellow to red. Only single pulses' bins where both $\sigma_{\text{P.A.}} \le 5^{\circ}$ and $\sigma_{\text{E.A.}} \le 5^{\circ}$ are included, to make sure that all presented points are with significant linear and circular polarization.\label{fig:PALI}}
		\end{figure}

\begin{figure}
			\centering
			\includegraphics[scale=0.28]{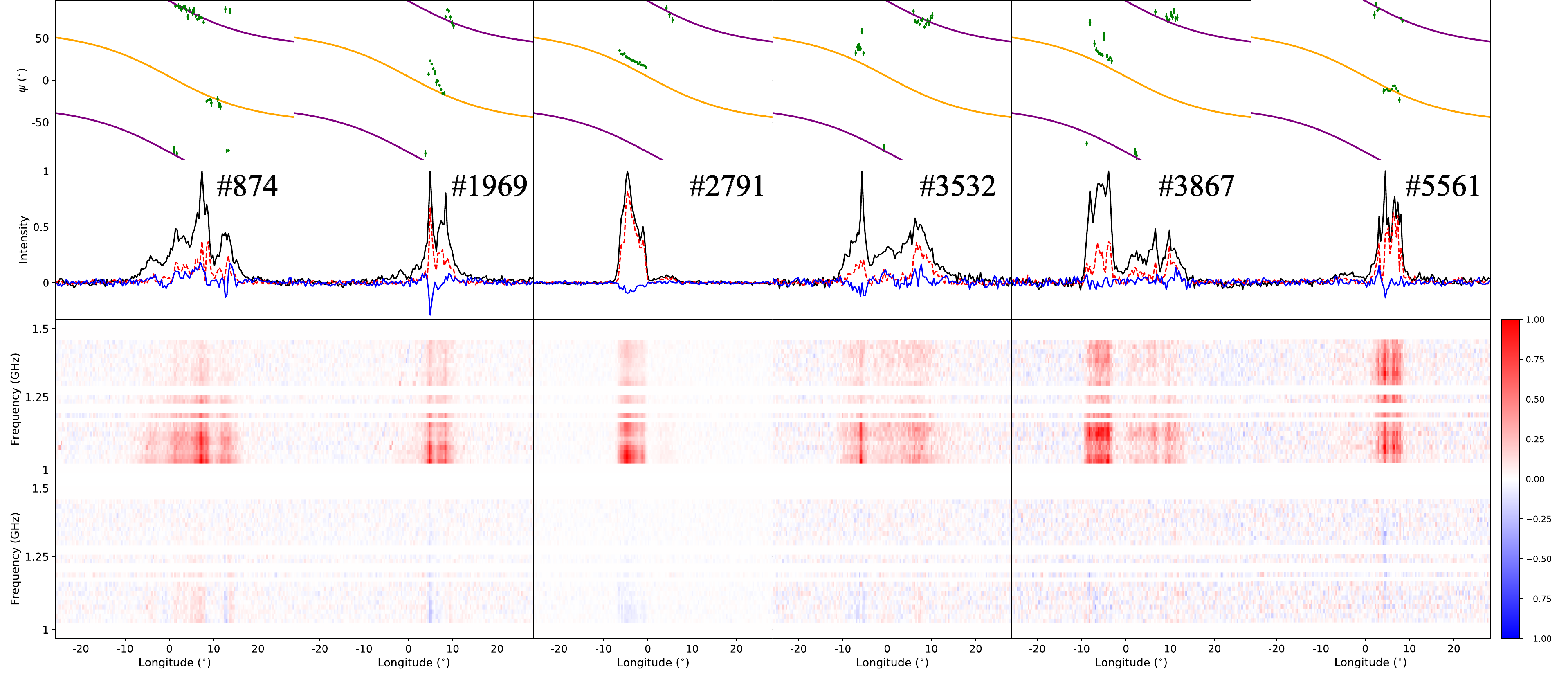}
			\caption{Examples of single pulses exhibiting orthogonal polarization modes. Panel 1 (1-4 from up to bottom): P.A. and RVM curves, where lines' and dots' meanings are same as Figure~\ref{fig:int_prof}. Here only P.A. points with $\sigma_{\text{P.A.}}\le 5^{\circ}$ are presented; Panel 2: individual pulse profiles of $I$, $L$ and $V$; Panel 3: dynamic spectrum of total intensity $I(\phi, \nu)$, where the 4096 frequency channels are binned into 32 bins (binned channels that are much affected by RFI are zapped). The intensity values are normalized with the maximum of $I(\phi, \nu)$; Panel 4: dynamic spectrum of circular polarization intensity $V(\phi, \nu)$, where the 4096 frequency channels are binned into 32 bins. The intensity values are normalized with the maximum of $I(\phi, \nu)$.\label{fig:single_z}}
		\end{figure}

		B0943$+$10's single pulses' distributions of $L/I$, P.A. and E.A. for all 4 epochs (14519 pulses in total) are shown in Figure~\ref{fig:PALI} (i). For the main pulse component longitude range ($-20^{\circ}$, $20^{\circ}$), there are P.A. patches located around $0^{\circ}$ and $\pm 90^{\circ}$, indicating the existence of OPMs. OPMs can also be seen directly from some individual pulses' P.A.s, like those shown in Figure~\ref{fig:single_z}. For the precursor component longitude range ($-70^{\circ}$, $-20^{\circ}$), there is only one P.A. patch, which is on the same RVM curve with the main pulse P.A. patch around $0^{\circ}$. The respective dominating polarization modes of the main pulse and the precursor are orthogonal, which is in accord with the P.A. curve fitting results in Section~\ref{sec:rvm}. The P.A., E.A. and $L/I$ distributions for B mode and Q mode are also shown in Figure~\ref{fig:PALI}(ii) (iii). In Figure~\ref{fig:PALI} (ii) for the B mode's case, the extra component at longitude $\phi \approx -15^{\circ}$ described in LRFS in Section~\ref{sec:230827} and~\ref{sec:220902} also clearly appears. The extra component tends to have pure SPM P.A. patch, higher $L/I$ and more negative $V$ than the main part of B mode pulse component.

\begin{figure}\centering
			
   \includegraphics[scale=0.27]{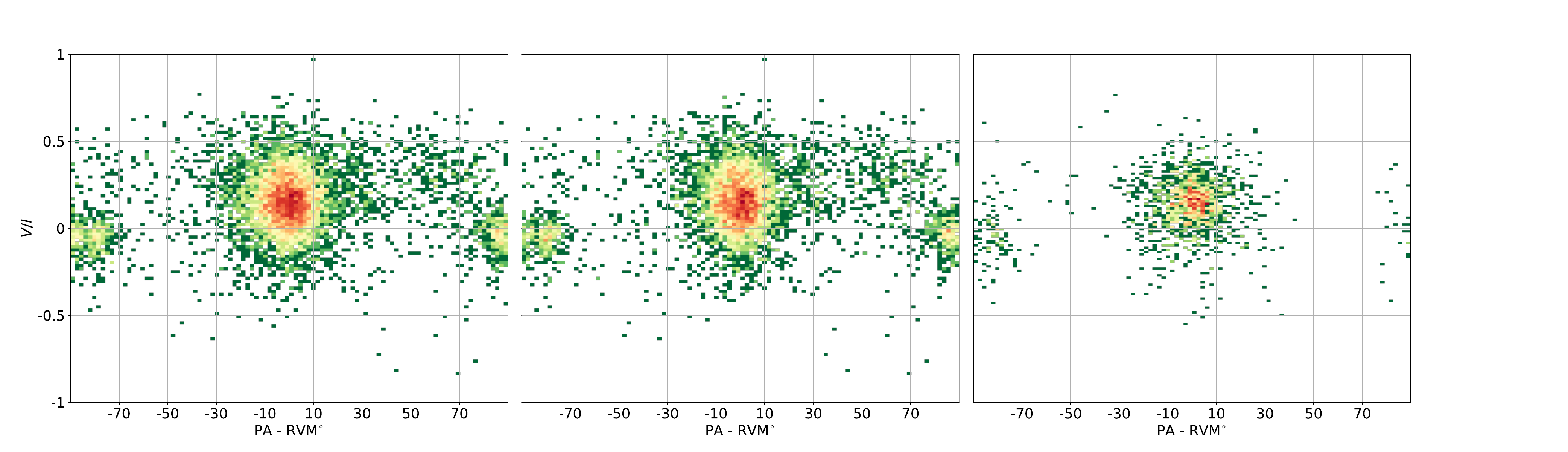}\centering
			\caption{P.A. - E.A. distribution for the main pulse component. Horizontal axis: P.A. subtracted by the purple RVM curve in Figure~\ref{fig:PALI}; vertical axis: E.A.. Only single pulses' bins where both the uncertainties $\sigma_{\text{P.A.}} \le 5^{\circ}$ and $\sigma_{\text{E.A.}} \le 5^{\circ}$ are included.\label{fig:PAEA}}
		\end{figure}

        It is worth noticing that although the integrated profile shows always $V>0$, a fraction of single pulses' ellipticity angles are actually below $0^{\circ}$, indicating some minus-sign $V$ appears significantly in single pulses. We wonder if different signs of Stokes $V$ correspond to different P.A. patches (or equivalently, different OPMs). Figure~\ref{fig:PAEA} shows the relation between circular polarization degrees $V/I$ and P.A.s subtracted by the purple RVM curve in Figure~\ref{fig:PALI}. Compared to Figure~\ref{fig:PALI} we find that pulse bins with P.A. around $\pm 90^{\circ}$ tend to have $V>0$. But for pulse bins with P.A. around $0^{\circ}$, they are more likely to have $V<0$. Moreover, the polarization mode with $V>0$ tends to have larger $|V|$. Compared to the results of~\cite{deshpande2001} (also introduced in Section~\ref{sec:intro}), in Figure~\ref{fig:PALI}, P.A. patch around $\pm 90^{\circ}$ is the primary polarization mode (PPM), and the P.A. patch around $0^{\circ}$ is the secondary polarization mode (SPM). The main pulse component has certain amounts of both PPM and SPM, but dominated by PPM. The precursor's P.A.s are on the same RVM curve with SPM patch of the main pulse component, indicating that the precursor component is pure SPM.
		
		In Figure~\ref{fig:int_prof}, the maximum linear polarization fraction, which could be affected by orthogonal polarization modes' incoherent mixing, doesn't vary evidently in main pulse components of B mode's and Q mode's integrated profiles. We'd like to examine if OPMs' proportions change between different modes at 1-1.5~GHz. In order to figure out PPM and SPM's proportion in B mode's and Q mode's main pulse component, we subtract the purple RVM curve in Figure~\ref{fig:int_prof} from the actual P.A. values again, to make SPM and PPM more concentrated around $\pm 90^{\circ}$ and $0^{\circ}$. The subtraction process is equivalent to calculating P.A. with respect to the purple RVM curve. For example, if an individual pulse's P.A. curve prefectly follows the RVM curve, then after subtraction, its P.A. curve becomes a horizontal line at $0^{\circ}$. 
        Regarding rotated P.A. points within range $(-45^{\circ}, +45^{\circ})$ as PPM P.A. points, others as SPM P.A. points, PPM and SPM pulse bins (with $\sigma_{\text{P.A.}} \le 5^{\circ}$ \& $\sigma_{\text{E.A.}} \le 5^{\circ}$) numbers in B mode's and Q mode's main pulse component phase range are counted and shown in Table~\ref{tab:Ortho_count}. The proportion of SPM versus PPM is larger for pulse bins with significant polarization in Q mode than that in B mode.

        \begin{table}
			\centering
			\renewcommand\arraystretch{1.2}
			\caption{MAIN PULSE ORTHOGONAL MODES COUNTING \label{tab:Ortho_count}}
			
			\begin{tabular}{ccc} 
				\hline
				\hline
				Radiation modes & Number of PPM & Number of SPM \\
				\hline & 
				PULSE BINS WITH SIGNIFICANT POLARIZATION ($\sigma_{\text{P.A.}} \le 5^{\circ}$ \& $\sigma_{\text{E.A.}} \le 5^{\circ}$) &\\ \hline
                B mode & 1453 & 144 \\
				Q mode & 4518 & 1126 \\
				\hline &
                FULL INDIVIDUAL PULSES & \\
                \hline
                B mode & 2607 & 879\\
				Q mode & 8364 & 873\\
				\hline
			\end{tabular}
		\end{table}

        However, a large fraction of pulse bins with low polarization flux are ignored if we only consider pulse bins with significant polarization. Those high-polarization-flux bins, with a number of 7241, are only very small fractions of all individual pulses' bins: each pulses are divided into 1024 bins, so there should be $0.2\times 1024\times 14521\approx3\times 10^6$ samples within the on-pulse longitude range. In order to take all pulses into account, we similarly subtract the purple RVM curve in Figure~\ref{fig:int_prof} from their $Q$ and $U$ points. Marking Stokes $Q$ and $U$ at single bin $\phi$ as $Q(\phi)$ and $U(\phi)$, and the orange RVM curve value at $\phi$ as $\delta(\phi)$, practically the subtraction could be denoted as:

        \begin{equation}
			\left(\begin{matrix}
				Q'(\phi) \\
				U'(\phi) \\
			\end{matrix}\right)=
			\left(\begin{matrix}
				\cos(2\delta(\phi)) & \sin(2\delta(\phi))\\
				-\sin(2\delta(\phi)) & \cos(2\delta(\phi))\\
			\end{matrix}\right)
			\left(\begin{matrix}
				Q(\phi) \\
				U(\phi) \\
			\end{matrix}\right)
		\end{equation}
        Then we calculate representative P.A.$^{*}$s for all individual pulses:
        \begin{equation}
            \text{P.A.}^{*} = 0.5\arctan(\dfrac{\Sigma_{\phi}U'(\phi)}{\Sigma_{\phi}Q'(\phi)})
        \end{equation}
        Although it has been mentioned above that both OPMs also appear in individual pulses (Figure~\ref{fig:single_z}), the P.A.$^{*}$ could reveal the dominating orthogonal mode in an individual pulse. Regarding pulses whose P.A.$^{*}$s within range $(-45^{\circ}, +45^{\circ})$ as PPM dominated pulses, others as SPM dominated pulses, the counting results are also shown in Table~\ref{tab:Ortho_count}. From Table~\ref{tab:Ortho_count}, we see that OPM proportions change vastly before and after mode switch. As for the question why maximum linear polarization fraction don't vary much between B and Q mode profiles, it may be due to the differences in PPM's and SPM's single pulse polarization degrees: PPM's polarization seems to be more significant than SPM (also from Table~\ref{tab:Ortho_count}).

  \section{Discussion}\label{sec:sec4}
		\subsection{The profile evolution}\label{sec:profile_evo}
		
		The possible existence of strange attractor pattern in pulsar mode switches (Figure~\ref{fig:decom_wr}) indicates that the pulsar dynamic system can be in chaos, switching between two or more states. The ``B' mode'', though happens after a mode switch from Q mode, is different from the transitive modes reported in~\cite{2023A&A...675A..87S}, because the ``B' mode'' lasts for a long time ($> 1000$ pulses).~\cite{2014ARep...58..796S} reports a continuous increase in B mode's flux, lasting several hours, after Q-to-B mode switch. So ``B' mode'' may be a beginning status of B mode.

		\subsection{The geometry and the polar cap region: towards understanding the X-ray emission}\label{sec:discussion_geo}

        \par Further analysis on the radiation geometry, the polar cap pattern, the emission height and the radiating particle's Lorentz factors in Section~\ref{sec:result_map} are carried out in this section. Firstly we'd like to point out that the RVM fitting result of B0943$+$10's FAST L band observations is different from the frequently mentioned radiation geometry in former studies, namely the geometry given by~\cite{deshpande2001}.~\cite{deshpande2001} calculates the geometry through radiation cone models (based on~\cite{1993ApJ...405..285R}) to fit for both polarization angles and pulse profiles, and their results are $\alpha\approx 12^{\circ}$ and $\beta\approx -5^{\circ}$ (see also Table~\ref{tab:R_PA}). 
        
        \par A question related to our mapping result in Section~\ref{sec:result_map} is the true origin of the radiating particles. A good RVM fitting indicates that at the particles radiating positions, the magnetic field is well described with a dipole field, which usually demands that the radiating position is not too far from the pulsar surface. And the existence of sub-pulse drifting in the main pulse region suggests that radiation originate further from the magnetic axis, according to the carousel model in \cite{1975ApJ...196...51R}. So from Figure~\ref{fig:geomapping}, the IAG might be a more reasonable birthplace of the radiating particles, for both the main pulse component and the precursor component. If we accept that statement, then comparing with the main pulse component, the precursor component is radiated higher from larger $\gamma$ radiating particles, as is already pointed out in Section~\ref{sec:result_map}.
		
		Based on the geometry derived from FAST observations in this paper, we could give explanations to the X-ray radiation properties of B0943$+$10. First on the X-ray pulsation observed in both B and Q mode, it has already been pointed out~\citep[e.g.,][]{2013Sci...339..436H,2019ApJ...872...15R} that the significant thermal X-ray pulsation could come from a local high-temperature spot on the pulsar surface, namely a hotspot.~\cite{2013Sci...339..436H} pointed out the possible contradiction between the observed X-ray pulsation and the radiation geometry given by~\cite{deshpande2001}: when the magnetic axis, the spin axis and the line of sight are almost aligned (with small $\alpha$ and $\beta$ angles), the hotspot in the polar cap region is more likely to appear unmodulately and thus X-ray pulsation will not happen. In our result's case, the axes are not so close to each other, making the hotspot modulation more feasible.
		
		However, there still exists a question why thermal X-ray pulsation is significant in Q mode, but almost ceases in B mode. X-ray usually has few interactions with magnetized plasma, so it is more possible that some changes on pulsar surface happening during mode switch lead to differences in hotspots of different radiation modes.
		
		A possible answer to this question lays in the origin of the precursor component. As is mentioned above, compared with the main pulse component, the precursor component is radiated higher (further from pulsar surface) by radiating particles with larger Lorentz factor $\gamma$. A simple picture on radiating particles \citep[e.g.,][]{1975ApJ...196...51R} is that all primary particles are accelerated to very large $\gamma_{\text{primary}}$ ($\sim 10^{6}$) near the pulsar surface, and then flow along magnetic field lines, losing energy through radiation, and high-energy photons produce secondary particles. After several turns of cascade reactions, finally radiating particles radiate radio frequency waves at certain heights, which is observed by us. According to estimations of primary particles' Lorentz factor like \cite{2023PhRvD.107h1301S}, we state that since the precursor component's radiating particles have larger Lorentz factors and has gone through longer ways along magnetic field lines losing more energy, the precursor component may originate from primary particles with larger $\gamma_{\text{primary}}$ than those contribute to the main pulse component.
		
		High-energy primary particles and secondary particles can flow back and hit the surface to form hotspots which lead to X-ray emission~\citep[e.g.,][]{2000ApJ...532.1150Z,2001ApJ...556..987H}. Under such logic, the component with higher-energy primary particles might be related with a hotter hotspot. In B0943$+$10's case, the hotspot associated with the precursor component is hotter than the hotspot associated with the main pulse component. It's also possible that the precursor component, or the Q mode profile as a whole, originates nearer to the magnetic axis than B mode (like the light purple dash-dotted line in Figure~\ref{fig:geomapping}), which results in higher energy secondary particles for Q mode radio radiation origin. All two processes above make Q mode X-ray pulsation stronger than B mode.

		The results in Figure~\ref{fig:geomapping} also provide possible explanations to some other phenomena observed. (1) Since the precursor component is radiated relatively closer to the light cylinder than the main component, the precursor radiation undergoes less propagation effects and less circular polarization is introduced by propagation, which could explain why the precursor component has high $L/I$. (2) At lower frequency bands, the curves in the ``Height'' panel of ~\ref{fig:geomapping} will have a overall shift to higher heights (according to radius-frequency-mapping). When the frequency is too low, the calculated precursor component's radiation height may exceed the light cylinder, which could cause the precursor component to actually be weaken, like Fig. 2 in \cite{2013Sci...339..436H}: comparing to the main pulse component, the precursor component is relatively much weaker under LOFAR 140~MHz observation than under Giant Metrewave Radio Telescope 320~MHz observation.

        We'd like to mention that in pulse profile plots with fitted RVM curves (Figure~\ref{fig:int_prof} and \ref{fig:geomapping}), the phase locations of main pulses' peaks are all close to longitude $\phi = 0^{\circ}$, namely the centroid of the RVM curve. If considering Blaskiewicz-Cordes-Wasserman (BCW) shift described in \citet{1991ApJ...370..643B}, the radiation height shown in Figure~\ref{fig:geomapping} (at $\phi = 0$, height $r \approx 300 R_{\text{NS}} \approx 3\times 10^{8}$cm) make the BCW shift be $\Delta \phi\approx 0.04$, which is larger than what we've observed. This may indicate that the birthplaces of radiation particles are actually closer to the feet of last open field lines, which means that radiation heights are actually lower (see Figure~\ref{fig:geomapping} (iii)'s black lines for the case of last open field lines). Besides, radio radiation's propagating through polarization limiting region will also reduce the BCW shift~\citep[e.g.,][]{1986ApJ...303..280B, 1991ApJ...370..643B}. Anyway, if BCW shift does work, the intensity profile's centroid will move to $\phi\approx -0.04$, which actually doesn't affect our main conclusion that the precursor component is radiated higher by charged particles with higher energy.
        
		\subsection{The OPMs and propagation effects in magnetosphere}\label{sec:opms_discussion}
		
		OPMs in pulsar are usually interpreted as O (ordinary) mode and X (extraordinary) mode for high frequency electromagnetic waves propagating in plasma. Theoretically, OPMs can be produced near the emitting point \citep{1979ApJ...229..348C} or at higher regions in magnetospheres through wave mode coupling~\citep{2001A&A...378..883P}. In Figure~\ref{fig:single_z} both orthogonal modes could be observed in one single pulse, and this may be caused by the asymmetry of particles' distribution in the magnetosphere: on some longitudes, O mode dominates X mode, or reversely.

        The OPM - $V$ relation revealed by Figure~\ref{fig:PAEA} (also seen in other pulsars like B2020$+$28~\citep{1978ApJ...223..961C}) could be a result of the changing of circular polarization during waves' propagating through the pulsar magnetosphere. Take radiative transfer of Stokes parameters in uniform media for example~\citep[e.g.,][]{1969SvA....13..396S,2011MNRAS.416.2574H}, the change of circular polarization $V$ could be written in the form:

        \begin{equation}
            \dfrac{dV}{ds}=-\eta_{V}I+\rho_{Q}U-\eta_{I}V
            \label{eq:VI}
        \end{equation}

        The $U$ is defined in the local coordinate system presented in \cite{2011MNRAS.410.1052S}. The coefficient $\rho_{Q}$ in Equation~\ref{eq:VI} measures the conversion between linear polarization and circular polarization (Faraday conversion or mode coupling~\cite[e.g.,][]{1998Ap&SS.262..379L, 2000A&A...355.1168P, 2011MNRAS.416.2574H}). For orthogonal modes, P.A.$_{O}=$P.A.$_{X}\pm90^{\circ}$, so the Stokes parameter $Q$ and $U$ have opposite signs for O mode and X mode at the same longitude. Then the circular polarization caused by conversion tends to be in opposite directions for OPMs, which is also consistent with calculations in \cite{2001A&A...378..883P}. The different $|V|/I$ of OPMs then could come from cyclotron absorption term $\eta_{V}I$, because Stokes parameter $I$ is always $>0$. Therefore, if simply take above arguments into account, we may get:

        \begin{equation}
            |(|V|/I)_{O} - (|V|/I)_{X}| \sim \int \eta_{V}ds
            \label{eq:VI_eta}
        \end{equation}

        From Figure~\ref{fig:PAEA}, the difference in $|V|/I$ is about 0.2 for OPMs in B0943$+$10. Of course Equation~\ref{eq:VI_eta} is too simplified because the pulsar magnetosphere is not uniform, and our knowledge on the absorption and conversion process is limited. But such kind of observational patterns do give us opportunities to quantitatively dig into the pulsar magnetosphere.
  
        As for the wave propagation process in the pulsar magnetosphere, a question is how to match PPM and SPM to the actual O and X modes. If adapting criterion given by~\cite{2010AstL...36..248A} and~\cite{2012MNRAS.425..814B}, where O-mode has minus sign of $(d\text{P.A.}/d\phi)\cdot V$ and X-mode has plus sign of $(d\text{P.A.}/d\phi)\cdot V$, then for the main pulse component, P.A. patches around $\pm 90^{\circ}$ (PPM) correspond to O-mode, while the P.A. patch around $0^{\circ}$ (SPM) corresponds to X-mode, and the precursor component consists of only X-mode. The switch from B to Q mode leads to the arise of X-mode dominated precursor, as well as the change of OPM proportions in the main pulse region. It's possible to give a picture of B-to-Q mode switch that something happening on the pulsar surface gives rise to quick changes of discharge regions in the polar cap and of some physical parameters (like the particle number density~\citep[e.g.,][]{2006Sci...312..549K}) of magnetosphere. Anyway, those physical processes may need more detailed investigations.

        \subsection{On the trigger of mode switches}
        From the analysis on geometry and single pulse polarization, the mode switch process seems to require some changes on pulsar surface. The surface condition's influence to pulsars' radio emission has been considered for the case of "mountains" or "zits" on the surface \citep[e.g.,][]{1981IAUS...95..173V,2024arXiv240114181W,Wang_2024}, and has been taken into account as an explanation to the mode switch phenomenon~\citep[][]{1982ApJ...258..776B}. The birth of a new spark discharge point may suppress spark discharging near it \citep{1982SvA....26..443B}, which may explain the main pulse component intensity's anti-correlation with the precursor component intensity pointed out in Section~\ref{sec:modeevo}. Meanwhile, it has been reported that planets are found around B0943$+$10~\citep{2014ARep...58..796S, 2019ARep...63..310S}.~\cite{2014ARep...58..796S} suggests that B0943$+$10's mode switch and X-ray emission are caused by surrounding matter's accretion. The accretion could lead to changes in magnetosphere and hotspots' formation on pulsar surface, too. To judge whether mode switch is triggered by surface change or external accretion or both, more observations on other wavelengths are needed.
		
		\section{Conclusion}\label{sec:sec5}
		
		PSR~B0943$+$10 is observed by FAST at 1-1.5~GHz. The radiation modes in all 4 observation epochs are analyzed. The mode switch process could be quantitatively described with the results of an eigen mode searching algorithm. The mixture weights of the two most significant eigen modes appear jumps that representing the mode switch process, and a strange-attractor-like pattern in 2D plane, suggesting the pulsar system switches between certain states. Under FAST L band, Q mode is almost as bright as B mode, indicating different frequency evolution between modes.
		
		Rotating vector model fitting of the integrated profile's polarization angles gives a geometry of $\alpha = 42^{\circ}$ ($138^{\circ}$) and $\beta = 14^{\circ}$ ($-14^{\circ}$). The precursor component and the main pulse component are orthogonally polarized. Based on this geometry we map the radiation to some representative magnetic field lines starting from the polar cap on the pulsar surface and estimate the radiation height and radiation particles' Lorentz factors. We conclude that B0943$+$10's L band radiation particles are more likely to originate from the annular region of the polar cap than from the core region. Comparing with the main pulse component radiation, B0943$+$10's precursor component radiation comes from a further place from the magnetic axis, has higher radiation height and larger Lorentz factors of radiation particles. So the precursor component may have primary particles with higher energy, leading to a hotter hotspot on pulsar surface. The precursor component only appears in Q mode, so B0943$+$10 may have considerable thermal X ray pulsation only in Q mode, which is consistent with former X-ray observations. 
		
		On the other hand, single pulses' polarization properties are studied. B0943$+$10's different orthogonal polarization modes (OPMs) tend to have different handedness of circular polarization, which could be used to relate observed OPMs to ordinary and extraordinary wave modes in plasma, as well as to give some possible constraints on the cyclotron absorption process in the magnetosphere. The proportion of OPMs changes with mode switches. Some individual pulses also appear with both two orthogonal modes. The precursor component's stimulation, as well as the change of polarization during mode switch, may be related to some variations on the pulsar surface.
		
		\acknowledgments
		We thank Dr. Lucy Oswald from Oxford University for participating in our observation proposal of epoch (b) epoch. We thank Dr. Abdujappar Rusul from Kashi University, Prof. Bing Zhang from University of Nevada, Prof. Jinlin Han from National Astronomical Observatories, Prof. V. S. Beskin from P. N. Lebedev Physical Institute and Prof. Sandro Mereghetti from Istituto di Astrofisica Spaziale e Fisica Cosmica for interesting discussions. An anonymous reviewer has provided us with many suggestions.
		All data used in this work is from the Five-hundred-meter Aperture Spherical radio Telescope (FAST). FAST is a Chinese national mega-science facility, operated by National Astronomical Observatories, Chinese Academy of Sciences. This work is supported by the National
        SKA Program of China (2020SKA0120100), the National Natural Science Foundation of China (Nos. 12003047, 11703047, 11773041, U2031119, 12133003, and 12173052), and the Strategic Priority Research Program of the Chinese Academy of Sciences (No. XDB0550300). Z. P. is supported by the Youth Innovation Promotion Association of CAS (id.
        Y2022027) the CAS ``Light of West China'' Program, and National Key R\&D Program of China, No. 2022YFC2205202. This work is supported by the National SKA Program of China (2020SKA0120100), the National Natural Science Foundation of China (Nos. 12003047 and 12133003), and the Strategic Priority Research Program of the Chinese Academy of Sciences (No. XDB0550300).

		\end{document}